\documentclass[twocolumn]{aastex62}
\pdfoutput=1 
\usepackage{amsmath,amstext}
\usepackage[T1]{fontenc}
\usepackage{apjfonts} 
\usepackage[figure,figure*]{hypcap}
\usepackage{graphicx}
\usepackage{rotating}
\usepackage[para,flushleft]{threeparttable} 
\usepackage{xcolor}

\shorttitle{X-ray Cepheids}
\shortauthors{S.~P. Moschou et al.}

\begin{document}

\title{Phase-modulated X-ray Emission from Cepheids due to Pulsation-Driven Shocks}

\author{Sofia-Paraskevi Moschou}
\affiliation{Center for Astrophysics | Harvard \& Smithsonian, 60 Garden Street, Cambridge MA02138, USA}

\author{Nektarios Vlahakis}
\affiliation{Section of Astrophysics, Astronomy and Mechanics, Department of Physics, National and Kapodistrian University of Athens, 15784 Zografos, Athens, Greece}

\author{Jeremy J. Drake}
\author{Nancy Remage Evans}
\affiliation{Center for Astrophysics | Harvard \& Smithsonian, 60 Garden Street, Cambridge MA02138, USA}

\author{Hilding R. Neilson}
\affiliation{University of Toronto, David A. Dunlap Department of Astronomy \& Astrophysics, 50 St. George Street, Toronto, Ontario,
Canada M5S 3H4}

\author{Joyce Ann Guzik}
\affiliation{Center for Theoretical Astrophysics, Los Alamos National Laboratory, XTD-NTA, MS T-082, Los Alamos, NM 87545, USA}

\author{John ZuHone}
\affiliation{Center for Astrophysics | Harvard \& Smithsonian, 60 Garden Street, Cambridge MA02138, USA}


\begin{abstract}
Cepheids are pulsating variable stars with a periodic chromospheric response at UV wavelengths close to their minimum radius phase. Recently, an X-ray variable signature was captured in observations during the \emph{maximum} radius phase. This X-ray emission came as a surprise and is not understood. In this work, we use the modern astrophysical code, PLUTO, to investigate the effects of pulsations on Cepheid X-ray emission. We run a number of hydrodynamic numerical simulations with a variety of initial and boundary conditions in order to explore the capability of shocks to produce the observed phase-dependent X-ray behavior. Finally we use the Simulated Observations of X-ray Sources (SOXS) package to create synthetic spectra for each simulation case and link our simulations to observables. We show that, for certain conditions, we can reproduce observed X-ray fluxes at phases 0.4--0.8 when the Cepheid is at maximum radius. Our results span a wide range of mass-loss rates,  $2\times10^{-13}$--$3\times10^{-8}$ $M_\odot$ yr$^{-1}$, and peak X-ray luminosities, $5\times10^{-17}$--$1.4\times10^{-12}$ erg cm$^{-2}$ s$^{-1}$. We conclude that Cepheids exhibit \textit{two component emission} with (a) shock waves being responsible for the phase dependent variable emission (phases 0.2 - 0.6), and (b) a separate quiescent mechanism being the dominant emission mechanism for the remaining phases.
\end{abstract}

\keywords{Stars: activity, flare, late-type, atmospheres, X-ray: stars, variable stars: Cepheids}


\section{Introduction}
\label{S:1}
Classical Cepheid variable stars are radially pulsating standard candles. They are important stars for understanding both stellar evolution \citep[e.g.,][]{Anderson.etal:16, Miller.etal:18} and for cosmological measurements \citep[e.g.,][]{Freedman.Madore:10,Ngeow.etal:15,Riess.etal:16}. This is because Cepheids are luminous yellow supergiants and their pulsation periods are correlated to their luminosities via the Leavitt Law \citep{Leavitt:1912}. This combination of properties allows us to understand these stars in great detail, but even after a century of the Leavitt Law, Cepheids are still offering new mysteries.

Even though the photospheric modulation in Cepheids has been studied extensively, little is known about the conditions higher in their atmospheres. Recently, \citet{Engle.etal:17} presented X-ray observations from both \textit{XMM-Newton} and \textit{Chandra} verifying the unusual X-ray behaviour of the prototype Cepheid, $\delta$ Cephei, that shows an enhancement by a factor 4 in its X-ray flux at phase 0.45, \citep[see Fig.~1][]{Engle.etal:17}. This is an indication that the enhancement is triggered by a pulsation mechanism.  The fractional X-ray luminosity of $\delta$~Cep is of the order $L_X/L_{bol}=10^{-8}$, i.e.\ $4\times 10^{28}$~erg~s$^{-1}$, with a peak luminosity of  $1.2\times 10^{29}$~erg~s$^{-1}$.
The case of $\delta$~Cep, however, is not unique. A handful of Cepheid observations have been made by {\it Chandra} revealing one more candidate that presents an increase in X-ray emission at the same phase as $\delta$ Cep, namely $\beta$ Doradus \citep{Evans.etal:18}. This X-ray behavior came as a surprise and is not currently understood. More recently during an \emph{XMM-Newton} follow up a low-mas companion was revealed from V473 Lyrae while the X-ray emission of V473 Lyr remained constant for the full time of the coverage corresponding to a third of the pulsation period \citep{Evans.etal:20}.


\citet{Ayres:17} and \citet{Ayres:18}  presented evidence that late-F and early-G supergiants, including Cepheid variables, might constitute a new class of coronal source, overluminous in X-rays and 10 times weaker in Si IV 1393 \AA\ chromospheric emission than later-type supergiants. A prominent member of this new class is the F5~1b star $\alpha$~Per. This supergiant appeared to be a significant coronal ($T = 10^7$ K) X-ray source, while having 10 times weaker Si IV 1393~\AA~($T = 8\times10^4$ K) emission from lower atmospheric layers than early-G supergiants of similar high-energy emission. The scenario of a secondary being the cause of this discrepancy was excluded by a follow-up multi-wavelength campaign. 
Shortly after, \citet{Ayres:18} presented additional evidence, from the observations of three early-F supergiants with Chandra in X-rays and HST in the far-ultraviolet (FUV), that early-F supergiants seem to belong to a distinct coronal class, characterized by elevated X-ray/FUV ratios. \citet{Ayres:18} suggested that this could imply thinner outer atmospheres on the F types, as in dwarfs, possibly due to a weakened ``ionization valve'' effect because of their hot photospheres.
These supergiants bear similarities to Cepheid variables in their X-ray enhancement phases and follow the same trend in X-ray/FUV ratio as low-activity late-type dwarfs.
While the X-ray emission in Cepheid during the enhancement phases bears similarities to the supergiant trend, the phase-dependent variability is still not understood.

One possible explanation for the enhancement of the X-ray signal is that it is due to plasma heated by shock waves triggered by the pulsation cycle. Another explanation, could be due to particle acceleration and energy release at magnetic reconnection regions.
While X-ray emission in normal G-K giants is probably of magnetic nature \citep{Auriere.eta:15}, in Cepheids, a hydrodynamic scenario that periodically produces shock waves is also possible \citep[e.g.][]{Sasselov.Lester:94,Engle.etal:14}. Outward moving acoustic waves could become shock waves due to the plasma density stratification in a Cepheid atmosphere \citep[e.g.][see also Fig.~\ref{fig:shocks}]{Sasselov.Lester:94}. According to perpendicular shock theory (see Fig.~\ref{fig:shocks}), a single shock wave will compress and heat the Cepheid's atmosphere leading to density and temperature enhancements up to factors of about 3 and 4, respectively. Such an energy release could explain the \emph{X-ray flux} increase by a factor 4 (intensity $I=f(n^2,T)$) observed by \citet{Engle.etal:17} for $\delta$~Cep. 

Cepheids and non-pulsating supergiant stars have similar FUV and X-ray emission fluxes \citep{Ayres:17}, so the pulsation might not be responsible for the quiescent X-ray emission. In this study, we explore whether pulsation- driven shock waves can reproduce the characteristics of the X-ray variable emission in $\delta$~Cep as observed and summarized in \citet{Engle.etal:17}. In general, Cepheids show weak X-ray emission, as discussed in \citet{Engle.etal:17}, with $F_X\le 10^{-14}$ erg s$^{-1}$ cm$^{-2}$. More specifically, as shown in Fig.~1 of \citet{Engle.etal:17}, the observed quiescent emission for $\delta$~Cep is  $F_X = 5\times10^{-15}$ erg s$^{-1}$ cm$^{-2}$ and it reaches at its peak a luminosity of $F_X = 20\times10^{-15}$ erg s$^{-1}$ cm$^{-2}$, i.e. a factor of 4 higher. Furthermore, \citet{Engle.etal:17} report a temperature of the order of 5--20 MK in the hot plasma at the X-ray maximum and suggest that shock-driven heating would require shock speeds larger than 200~km~s$^{-1}$ to produce the observed X-ray emission. Finally, the increase in the X-ray light curve of $\delta$~Cep starts at phase 0.4, reaches its peak at 0.5 and drops back to quiescent levels around phase 0.6. So the X-ray enhancement appears to last in phase for a duration of 0.2. 

Another open question for Cepheids is their mass-loss rates. Several studies using both observations and theory have  estimated the mass-loss of $\delta$~Cep and other pulsating stars, with a wide range of mass-loss rates reported, ranging from $10^{-10}$ to $10^{-6}$ \citep[see, e.g.][]{Neilson.Lester:08,Marengo.etal:10b,Neilson.etal:11,Matthews.etal:12}. The radial velocity of $\delta$~Cep is a periodic function with an amplitude of about $18$ km s$^{-1}$ and an average value corresponding to the relative speed with respect to the Earth $16.4$ km s$^{-1}$. Keeping in mind these constraints, in the present work, we will be focusing on whether shocks are able to reproduce the mass-loss and X-ray luminosity values reported in literature so far. 

The paper is structured as follows. Section~\ref{sec:scenarios} introduces and discusses shocks in the atmospheres of pulsating stars, while Section~\ref{sec:normalshock} introduces the normal shock theory we will exploit in this study. Extensive numerical simulations are then described in detail in Section~\ref{sec:numericalsims}.
Although we use $\delta$~Cep as our guiding case for the mass and pulsation period of the star, we perform a number of experiments that can help us understand other cases of classical Cepheids with different observational signatures. This is discussed in greater detail in the context of the extant observations in Section~\ref{sec:discussion}, and the  summary and conclusions of the study are presented in Section~\ref{sec:conclusions}.

\section{Shocks in Atmospheres of Pulsating Stars}
\label{sec:scenarios}

The main difficulty in modeling pulsation-driven shocks is that the pulsation process is a very dynamical mechanism, that requires modelling the atmosphere and envelop of the star spanning orders-of-magnitude changes of temperature, density and pressure. Thus, although much progress has been made in modeling stellar atmospheres of pulsating stars, due to the complexity of the problem most models had to (a) focus on sub-photospheric up to low atmospheric regions, (b) make significant simplifications. The presence of multiple shocks can be seen in radial velocity observations of several RR Lyrae pulsating stars as localised bumps and bulges appearing in different parts of the pulsation cycles, and thus pointing towards strong acceleration processes due to shocks in the low atmospheres \citep[see, e.g.][]{Chadid.Preston:13}. Two shocks were found to occur per pulsation cycle in the special case of X Sgr Cepheid \citep{Mathias.etal:06}. Photospheric shocks could then accelerate as they propagate through a stratified atmosphere and potentially cause the X-ray enhancements observed, e.g. for the case of $\delta$~Cep \citep{Engle.etal:14}. 

 As shown in \citet{Kraft:60}, the H$_\alpha$ line is very sensitive to local conditions for both absorption and emission. \citet{Sasselov.Lester:90} produced velocity profiles for several Cepheids and found a phase-dependent absorption component for stars such as $\delta$~Cep. More recently, \citet{Nardetto.etal:08} studied Cepheids with both short- and long-periods and showed that H$_\alpha$ profiles present strong asymmetries that can only be explained by shocks. \citet{Nardetto.etal:08} also showed that in long-period Cepheids the observed H$_\alpha$ emission line can be caused by a shock wave propagating in the upper stellar atmospheric layers. 
 \citet{Wallerstein.etal:15} presented new radial velocity curves for $\delta$~Cep from a couple of dozen lines, as shown in Table~2 therein. H$_\alpha$ showed the highest velocity amplitude. In the case of $\delta$~Cep, H$_\alpha$ is only seen in absorption as the photospheric shock is not sufficiently strong, as explained in \citet{Gillet.Fokin:14}.
 
 \citet{Hill:72} was the first one to include in his models a large enough extent of the atmosphere to allow for the full development of shocks. He showed that the outer atmospheric layers are desynchronized with respect to the photosphere, as the atmosphere grows by 21 times during a pulsation cycle compared to the static case. Later on \citet{Fokin:92} took these kinds of models a step further, by considering, instead of the standard piston driver, the $\kappa$ mechanism within the H and He ionization zones as the pulsation driver.

Both \citet{Hill:72} and \citet{Fokin:92} found that there are two strong shocks forming in each pulsation period. The first or ''main'' shock forms during the atmospheric expansion phase and the secondary shock during the contraction phase.  Later on, \citet{Fokin.Gillet:97} in their pure radiative models obtained five shocks per pulsation cycle, i.e. the main and secondary waves, and another three shocks with lower amplitudes. When the same model was applied in the case of $\delta$~Cep, \citet{Fokin.etal:96} found that 
there are only four shocks forming, all with significant amplitudes, in each pulsation cycle. Then,  \citet{Fokin.Gillet:97} showed that for RR~Lyrae all these waves merge higher up in the atmosphere and produce the main shock that causes the H$_\alpha$ emission, with the highest intensity around phase 0.9 of the pulsation cycle. 

 Emission lines due to radiative shocks in pulsating stars, and specifically RR Lyrae, have been extensively examined \citep{Gillet.Fokin:14,Gillet.etal:17,Gillet.etal:19}.
 \citet{Gillet.Fokin:14} found that the Doppler shift for Balmer emission lines is about one third of the shock front velocity, $V_{shock}$, in the frame of the observer. However, hydrogen emission lines are not observed in classical Cepheids, as the atmospheric absorption completely dominates. As suggested by \citet{Gillet.Fokin:14}, this is an indication that full development of the main shock takes place far away from the photosphere. The authors further explain that the absence of Balmer emission lines in almost all short-period (less than 10 days) Cepheids observed are due to the fact that, even though the shock already forms at the photospheric level, initially it only has a very low Mach number and thus cannot cause notable emission.
 As further argued by \citet{Gillet.Fokin:14} using the Rankine-Hugoniot energy equation and assuming that the shock is very strong and thus can be considered isothermal, the radiative flux, $F_{rad}$, due to the shock is $F_{rad}\leq \frac{1}{2}\rho_0 V_{shock}^3$.

\citet{Sasselov.Lester:94} used non-local thermodynamic equilibrium radiative transfer models with hydrodynamics to study the chromospheric structure of classical Cepheids. They concluded that Cepheids have chromospheres heated by wave dissipation (acoustic or magnetic) on top of any pulsation-driven shock wave heating. \citet{Sasselov.Lester:94} found three shocks propagating in the pulsating Cepheid atmosphere and accelerating exponentially with steepening shock fronts as they move away from the chromosphere and into the transition region.
By observing and modeling the He~I~$\lambda$10830 line  \citet{Sasselov.Lester:94} also found a steady material outflow in upper Cepheid chromospheres and concluded that Cepheids are unlikely to sustain hot coronae.   
This view changed, however, once observations with HST, \textit{XMM Newton} and \textit{Chandra} telescopes became available. 

Later on, \citet{Engle.Guinan:12} presented an overview of X-ray, FUV and optical observations of classical Cepheids. They discussed several emission lines observed with \textit{HST/COS} corresponding to plasmas with temperatures $10^4$ up to $3\times10^5$~K, and X-ray observations with \textit{XMM Newton} and \textit{Chandra} corresponding to hot plasma of the order of a few $10^6$~K. For Polaris, the coolest line O I 1358~\AA\ (30,000 K) appears to be peaking first around phase 0.86, followed by Si IV 1400~\AA\ (100,000 K) and eventually N V 1240~\AA\ (200,000 K) \citep{Engle.Guinan:12}.

Recently, \citet{Boulangier.etal:19} simulated the stellar wind from slowly rotating pulsating stars on the Asymptotic Giant Branch (AGB) using a self-consistent hydrochemical model. \citet{Boulangier.etal:19} focused on very long period pulsations of the order of 300 days, with $M_\star = M_\odot$, $R_\star= 215 R_\odot=1$ AU, $T_\star=2500$~K $\rho_{\star}=10^{-9}$ g cm$^{-3}$($\approx 6\times10^{14}$ cm$^{-3}$), with a very steep density decay power law $\rho(r)=\rho_\star(r/R_\star)^{-10}$, and a velocity at the inner boundary given by a sinusoidal function with amplitude 2.5 km s$^{-1}$ up to 20 km s$^{-1}$. However, they did not account for the effects of magnetic fields and dust. They found that, in the purely hydrodynamic (HD) case, no sustained wind could be formed by  driving with observed pulsation amplitudes, while the temperatures in the simulation were too high for the formation of dust grains. \citet{Boulangier.etal:19} concluded that in the hydrochemical case sustaining a wind was even more difficult due to the efficient cooling mechanism. Nevertheless, in regions with high density, temperatures were low enough for dust to start forming. Naturally, such regions are close to the star, which is necessary for the stellar radiative pressure on dust grains to drive a sustained wind-like outflow.



%


\section{Normal Shock Theory}
\label{sec:normalshock}

\label{ss:normalshock}
\begin{figure}[htbp]
\begin{center}
\includegraphics[width=0.45\textwidth]{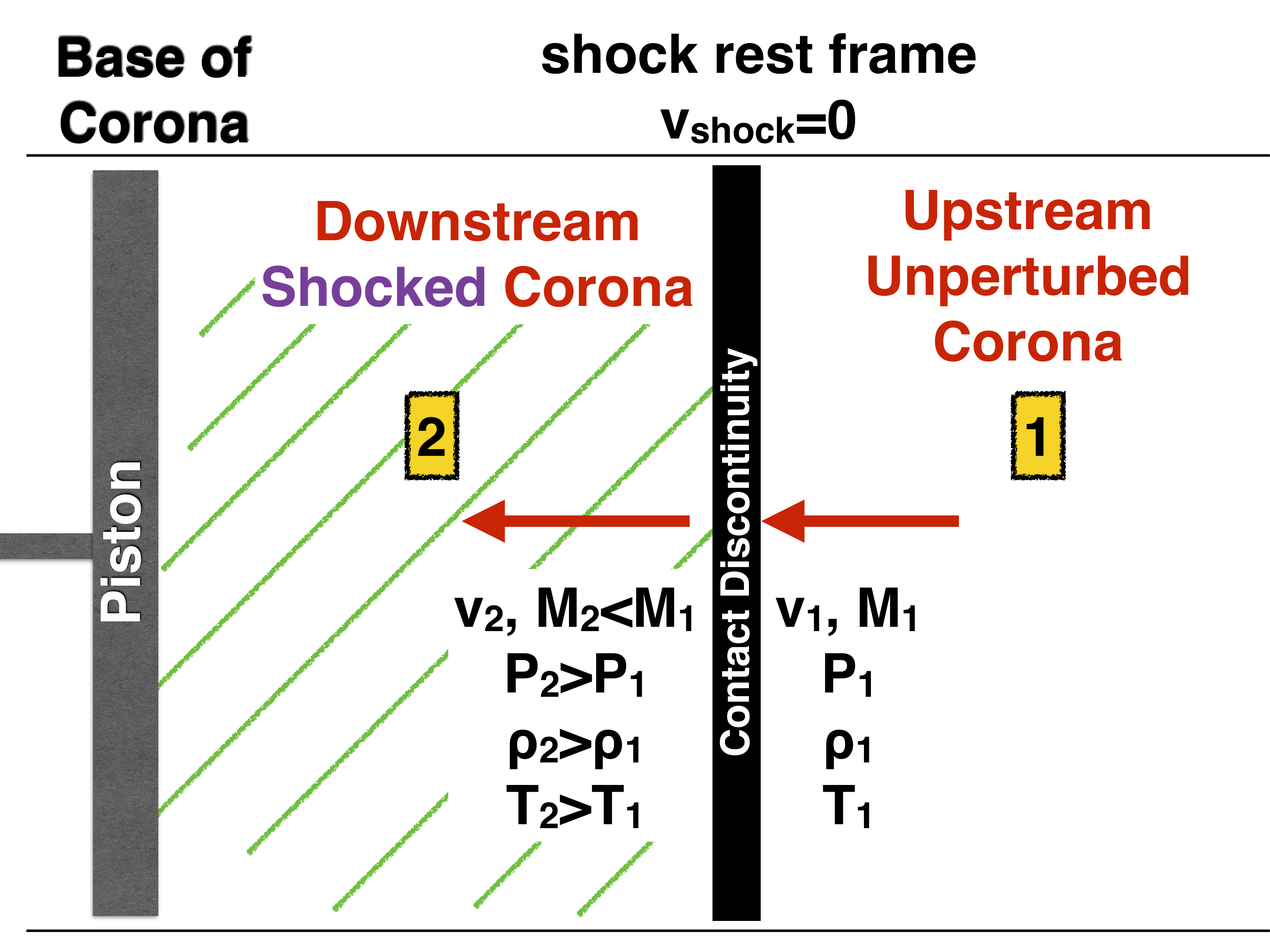}
\includegraphics[width=0.45\textwidth]{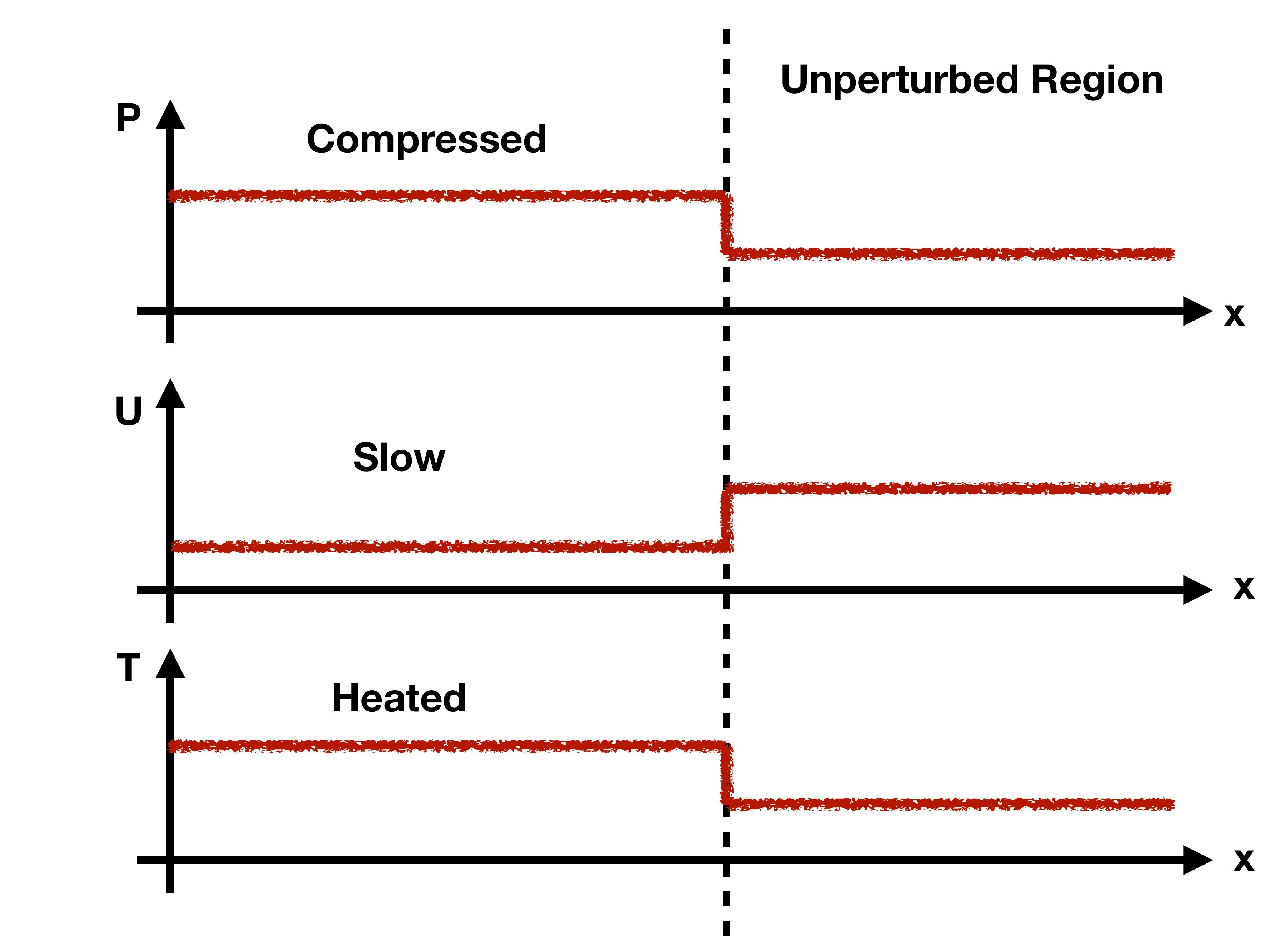}
\caption{Schematic of the shocked and unperturbed corona due to a normal shock wave triggered by pulsation in the shock's reference frame (M: Mach number).} 
\label{fig:shocks}
\end{center}
\end{figure}

Fig.~\ref{fig:shocks} demonstrates an example for the creation of a discontinuity that evolves into a shock. Let us assume that we have a long cylindrical container filled with a mono-atomic gas (polytropic index $\gamma=5/3$), which is closed on one side with a moving piston. Assuming that the piston moves with a constant speed $\mathrm{u_0}$, it will cause an increase of density and pressure. The information of that density and pressure increase will be transmitted throughout the entire cylindrical container with the speed of sound $\mathrm{c_s}$. As explained in the book by Landau and Lifshitz \citet{landau2013fluid}, the maximum speed for a steady flow is $u_{max}=c_s\sqrt{2/(\gamma-1)}\approx 1.7c_s$ for $\gamma=5/3$. If the speed of the piston is larger than the speed of sound locally $\mathrm{u_0}>\mathrm{c_s}$, there is not enough time for the atoms to transmit the information of the perturbation smoothly, since the sound waves are slower than the piston. The fluid solves this communication problem by increasing the sound speed $\mathrm{c_s}$ locally to a value higher than the speed of the piston with a subsequent increase in the temperature, pressure and density. The piston pushes the fluid violently and a discontinuity that moves with a sound speed larger than the piston speed $\mathrm{c_s}>\mathrm{u_0}$ forms. The hydrodynamic quantities at the fluids left (2) and right (1) of the discontinuity are linked through the Rankine \-\ Hugoniot jump conditions, which for the normal shock case can be written as follows in the shock's system of reference:
\begin{eqnarray}
\label{eq:rr}
    rr = \frac{\rho_2}{\rho_1}=\frac{u_1}{u_2}=\frac{\gamma+1}{\gamma-1+2/M_1^2}, \\
\label{eq:pp}
    \frac{P_2}{P_1}=\frac{2\gamma M_1^2-\gamma+1}{\gamma+1},
\end{eqnarray}
where $rr$ is the compression ratio, $\rho$ the mass density and the Mach number $M_1$ and the sound speed in fluid (1) are:
\begin{equation}
    M_1=\frac{u_1}{c_1}, \quad c_1=\sqrt{\gamma\frac{P_1}{\rho_1}}.
\end{equation}
Then combining Eq. (\ref{eq:rr}) with (\ref{eq:pp}) we get
\begin{equation}
\label{eq:pp2}
      \frac{P_2}{P_1}=\frac{(\gamma+1)rr-(\gamma-1)}{(\gamma+1)-(\gamma-1)rr},
\end{equation}
with the maximum compression ratio for a shock wave determined by the polytropic index $1<rr<\frac{\gamma+1}{\gamma-1}$ for $\gamma=5/3$ $1<rr<4$.

If the enhancement by a factor 4 in the X-ray signal observed by \citet{Engle.etal:17}, for $\delta$ Cep  $I^\prime=4I=f(n^2,T)$, is due to heated plasma from shock waves caused by the pulsation cycle, we need either a temperature and/or density increase that reproduces the observations. Thus, if the observed emission change is attributed purely to a temperature increase, we can find all the jump conditions of the shock. For this exercise, we use an approximation of the cooling function as a function of temperature in the range of temperatures of interest, i.e. between 10$^5$~K and a few 10$^5$~K.

\begin{figure}[htbp]
 \begin{center}
 \includegraphics[width=\columnwidth,trim={0cm 0cm 0cm 0cm},clip]{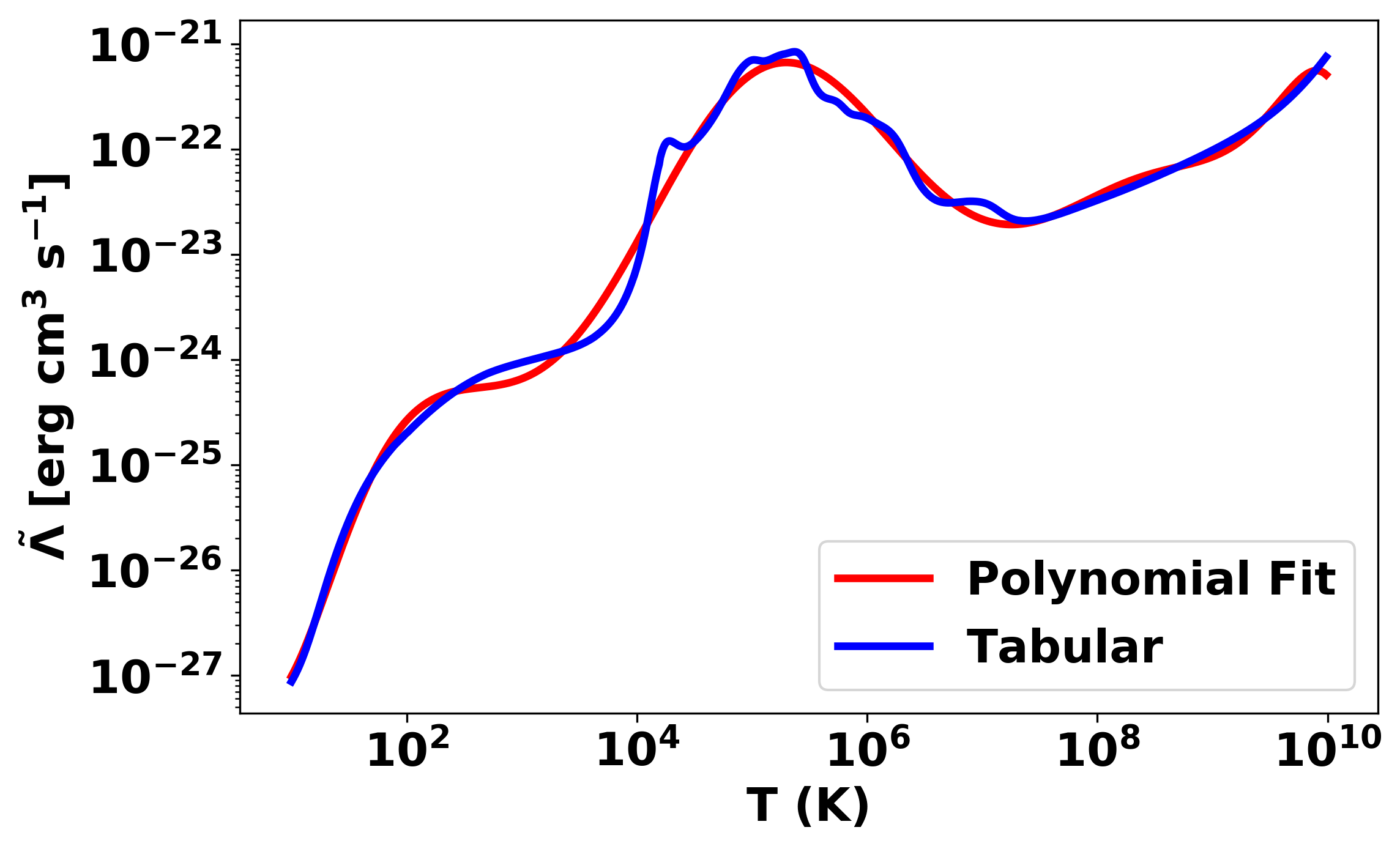}
  \caption{Cooling function obtained by fitting cooling table. Blue curve: tabulated cooling function, Red Curve: fitting function.}
  \label{fig:lambda}
  \end{center}
  \end{figure}

Optically-thin radiative losses, $\Lambda$, are generally a function of density and temperature $\Lambda = n^2 \Tilde{\Lambda}(T)$, with $n=\rho/{\mu m_u}$, and $\mu$ the mean molecular weight. The cooling function $\Tilde{\Lambda}(T)$ as a function of temperature is generally not known analytically, but there have been efforts to calculate it in the form of a table for solar coronal plasma \citep{Landi.Landini:99,Colgan.etal:08}, see also Fig.~\ref{fig:lambda} later in this manuscript. The cooling time can be calculated as 
\begin{equation}
\label{eq:lambda}
    \Lambda \tau_c = n k_B T \Rightarrow \tau_c =\frac{k_B T}{\Tilde{\Lambda}(T) n},
\end{equation}
where $\tau_c$ is the characteristic cooling time and $k_B$ is the Boltzmann constant.
Next, we use the tabulated cooling function shown in Figure~\ref{fig:lambda} to calculate the jump conditions of a single shock wave that matches the observed increase in the X-ray emission for different cases.

 A. Assuming the emission is thermal Bremsstrahlung for temperatures T$>$10$^7$~K, i.e. the emitted power per cubic centimeter is $e_{ff}=1.4\times10^{-27}T^{1/2}n^2 Z^2g_{Gaunt}$~erg/cm$^3$/s, and attributing the emission increase purely to a temperature increase, we have for the shocked (fluid [2] in Fig. \ref{fig:shocks}) to unperturbed (fluid [1] in Fig. \ref{fig:shocks}) HD ratios:
 \begin{eqnarray*}
 T_2/T_1=16,\quad P_2/P_1=60,\quad \rho_2/\rho_1=3.8, \\
 M_2=0.46,\quad M_1=7 
 \end{eqnarray*}
 This implies a very strong shock wave. 
 From the above and the continuity equation, $\rho_1\cdot v_1=\rho_2\cdot v_2$, we have $v_1=\rho_2/\rho_1\cdot v_2 \Rightarrow v_1 \approx 4 v_2 $.
 If we attribute the observed emission change purely to a density increase, we have for the shocked (fluid [2] in Fig. \ref{fig:shocks}) to unperturbed (fluid [1] in Fig. \ref{fig:shocks}) HD ratios:
 \begin{eqnarray*}
 T_2/T_1=1.75,\quad P_2/P_1=3.5,\quad \rho_2/\rho_1=2.0, \\
 M_2=0.65,\quad M_1=1.73 
 \end{eqnarray*}
 From the above and the continuity equation: $\rho_1\cdot v_1=\rho_2\cdot v_2$, we have $v_1=\rho_2/\rho_1\cdot v_2 \Rightarrow v_1 =2 v_2 $. 
 The above is just a mental experiment to gain some insights on the different regimes and the density and temperature ratios expected in each case. In reality, both temperature and density contribute to the emission, and for a shock wave, as Eqs. (\ref{eq:rr}) - (\ref{eq:pp2}) indicate, a shocked medium will always be heated.

 B. Assuming that the cooling function can be approximated by $e_{ff}\propto T^{-\alpha}n^2$~erg/cm$^3$/s, with $\alpha=-0.77$ for temperatures $2\times10^5<$T$<2\times10^7$~K, we have T$_2/$T$_1=0.165$. A temperature ratio lower than unity indicates that a rarefaction wave that cools the plasma down can also cause the increase in the emission observed, if the emission modulation is fully attributed to a temperature difference. However, this case is simply impossible, because a rarefaction wave would mean a compression ratio less than unity, according to the jump conditions described by Equations (\ref{eq:rr}) - (\ref{eq:pp2}). More importantly the density square dependence of the emissivity means that a temperature change in the range  $2\times10^5<$T$<2\times10^7$~K can never fully cause the observed emission increase, as the density contribution is more important. In fact, with the assumed emission spectrum, there is no combination of temperature and density ratios that satisfies the jump conditions and allows for an emissivity increase of a factor 4.

Stellar atmospheres, and especially Cepheid atmospheres, are very complex inhomogeneous regions, and the simple analytic or semi-analytic calculations that were presented so far do not suffice to accurately answer the question of whether shock waves triggered from the pulsation of the Cepheid can explain the periodic X-ray emission observed. After maximum radius the photosphere is beginning to contract. Therefore an outward propagating shock is moving against a flow with inward velocity and a velocity gradient. These colliding flows lead to shock formation that could be accelerated and have more energy than in the static case. For that reason, we will continue this analysis employing HD simulations. We start from idealised one-dimensional hydrodynamic simulations in order to reproduce the basic observational Cepheid behavior and keep adding physical ingredients to examine the $\delta$~Cep case as realistically as possible.



\section{Numerical Simulations}
\label{sec:numericalsims}

While the analytic intuition that the normal shock theory offers is valuable in understanding shock wave dynamics, in order to properly capture the dynamics of pulsation-driven shocks in Cepheid atmospheres we need to utilize self-consistent hydrodynamic simulations combined with radiative transfer considerations.

For our simulations, we will use the astrophysical fluid code \emph{PLUTO}, which provides a complete set of tools for this investigation and is described in \citep{Mignone:14}.
The advantages of using \emph{PLUTO} to model different
Cepheid atmospheric activity scenarios are that 
a) \emph{PLUTO} is a versatile code providing a multi-physics, multi-algorithm modular environment particularly suitable for modeling astrophysical flows in the presence of discontinuities \citep{Mignone.etal:07,Mignone:14} in any geometry (Cartesian, Spherical, Cylindrical);
b) it is one of the most user-friendly plasma codes with a simple user-interface based on Python, which is available to setup a physical problem in a quick and self-explanatory way;
c) it offers a collection of non-ideal energy source (and sink) terms, such as \textit{radiative cooling} and \textit{heating}, and diffusive terms, such as viscosity and resistivity, that are already prescribed and validated \citep{Mignone.etal:07,Mignone:14};
d) it allows for use of Adaptive Mesh Refined (AMR) grids;
e) it allows for data-driven runs and inclusion of observed magnetic fields that can have a dramatic effect on Cepheid wind structure, and it has already been used to model stellar winds of Sun-like and more active stars \citep{Reville.etal:16};
f) it allows for 1D to 2D, and 3D generalizations of the same physical problem in a straightforward manner compared to other "black box" codes specialized for specific physical scenarios and which do not allow for the same flexibility. This last point is essential, since in order to guarantee success and full control over the results we will reproduce the observed pulsation and wind characteristics in 1D before adding more dimensions and thus complexity in the problem. Magnetic reconnection cannot be studied in 1D simulations; at least 2D are required.

Pulsation-driven shocks in the stellar atmosphere can be modelled with non-resistive (ideal) HD.
The physical ingredients that need to be accounted for in the study of X-ray variability in Cepheid atmospheres include: 
1) a global stratified atmosphere, i.e. include the gravitational force of the Cepheid in a spherical geometry in order to capture the inverse square decrease of the density with distance, 
2) an initial sub-sonic hydrodynamic outflow, \citep[see, e.g.][]{Reville.etal:16}, 
3) a prescribed pulsation modulation \citep{Sasselov.Lester:94}, that after a large number of pulsation cycles will result in an atmospheric model in hydrodynamic equilibrium, and 
4) optically-thin radiative losses \citep{Mignone:14}. 

In this study, we will explore the shock wave scenario focusing on the case of $\delta$ Cep, in order to determine quantifiable emission characteristics of the shock-wave heating scenario and understand past and design future {\it Chandra} observations.

\subsection{Numerical Method}\label{sec:methods}

Radiative pressure in Cepheid atmospheres is several orders of magnitude lower than the dynamic gas pressure, according to photospheric models by \citet{Lester.Neilson:08} using the \emph{SATLAS} code. More specifically, \citet{Lester.Neilson:08} showed that the gas pressure is 3 to 4 orders of magnitude larger than the radiative pressure, and that at the expansion phase the electron number density is of the order of $10^7$ cm$^{-3}$, and the sound speed reaches 30 km s$^{-1}$. Thus, we can consider radiative pressure negligible and ignore it for the purposes of this study. 
In this paper, 
We then use \emph{PLUTO} to solve the following hydrodynamic set of equations with an ideal gas equation of state:
\begin{eqnarray}
\frac{\partial \rho}{\partial t}+\mathbf{u}\cdot\nabla\rho+\rho\nabla\cdot\mathbf{u} = 0 \\
\frac{\partial \mathbf{u}}{\partial t}+\mathbf{u}\cdot\nabla\mathbf{u}+\frac{\nabla p}{\rho} = \mathbf{g} \\
\frac{\partial p}{\partial t}+\mathbf{u}\cdot\nabla p+\rho c_s^2\nabla\cdot\mathbf{u} = 0.
\end{eqnarray}
For ideal gases the ratio of specific heats $\gamma$ is constant and the internal energy $e$ can be written as:
\begin{equation}
\rho e =\frac{p}{\gamma-1}.
\end{equation}

In short, we will use physical arguments to constrain the observed pulsation behaviour of $\delta$ Cepheid \citep[similar to][]{Sasselov.Lester:94} and explore possible HD scenarios for heating deposition in the Cepheid corona. We will start from the initial and boundary conditions described in \citet{Boulangier.etal:19} to reproduce the long period pulsation and wind profile of $\delta$~Cep similar to Fig.~\ref{fig:shocks} (bottom panels). 
The final ingredient will be to account for optically-thin radiative losses for the X-ray decay, which is a readily available module in \emph{PLUTO} \citep{Tesileanu.etal:08}. 

\paragraph{{\bf $\delta$~Cep Characteristics:}} The pulsation period of $\delta$~Cep is $T_p=5.366$~d \citep{Engle.etal:17}. It is \emph{approximately} five times more massive than the Sun, $M=5M_\odot$ {\citep{Caputo.etal:2005,Anderson.etal:2015,Guzik.etal:2020}}, and forty times larger, $R=40 R_\odot$. As shown in the radial velocity (fifth) panel of Fig. 1 of \citet{Engle.etal:17}, $\delta$~Cep moves towards the Earth with a speed of 16.4~km s$^{-1}$ and pulsates with a velocity amplitude of the order of 18~km s$^{-1}$. The escape speed of $\delta$~Cep at the photosphere is 219~km s$^{-1}$.

\paragraph{{\bf Units:}}
\emph{PLUTO} works with non-dimensional ``code'' units, and to avoid occurrences of extremely small or large numbers that complicate the calculations, we chose to scale the simulation setup to characteristic scales of the system. We can scale the results by specifying three basic units. The length unit is taken as the radius of $\delta$~Cep $L_0=40R_\odot=2.8\times10^{12}$~cm, the density unit is derived from the radiative cooling timescale  (see Section~\ref{sec:times} and Fig.~\ref{fig:times}) as $\rho_0=1.67\times10^{-15}$~g~cm$^{-3}$ or $n_e=10^{9}$~cm$^{-3}$, and the velocity unit is taken as $u_0=10^7$~cm~s$^{-1}$. The calculations presented in Section~\ref{sec:times} and Fig.~\ref{fig:times} are performed based on the assumption of a static, fully shocked atmosphere and provide only an order of magnitude estimation for the HD quantities that allow the production of the observed X-rays.

\paragraph{{\bf Grid:}} \emph{PLUTO} allows for different grid geometries. For this study we use exclusively spherical setups, as we can keep the numerical setup 1D and while still gaining insights for the 3D problem, since in spherical coordinates we properly capture the radial profile of the density as an inverse square law $\propto r^{-2}$ with distance $r$ from the star. 
The Cepheid atmosphere is a complex and extended region. In order to capture the dynamics in the lower atmosphere to high accuracy and at the same time propagate the solution through hundreds of astronomical units (AU), for the majority of the simulations presented here, we use 
a linear grid of 496 points in the region [1,50] $R_{\delta-Cep}$ and a logarithmically spaced grid of 100 points in the region [50,500] $R_{\delta-Cep}$.

\paragraph{{\bf Initial Conditions:}} Regarding the initial temperature for the Cepheid atmosphere, we assume coronal temperatures similar to those observed for the solar corona. Cepheids were considered to have cool atmospheres. However, recently \citet{Ayres:17} showed that F-G supergiants have hot coronae based on X-ray and FUV observations. 


For all our simulations we use a polytropic index of state 5/3. We choose two possible initial temperatures for the corona of $\delta$~Cep: $T=2\times10^6$~K corresponding to the temperature of the solar corona, and $T=5\times10^5$~K corresponding to a cooler corona.  A cooler corona might expected on the basis of a slowly-rotating low activity star assuming that magnetic fields are responsible for heating \citep[see, e.g.][for Alfv\'{e}n wave dissipation heating]{Cranmer.Saar:11}. 
In order to obtain a profile for the stratified atmosphere, we use an initial radial profile for the density corresponding to a steady outflow at constant velocity defined as $\rho = \rho_0 R_\star/r^2$, where $\rho_0$ is the initial density at the inner boundary, as shown in Table~\ref{tab:sims}.
Then, the radial profile of the pressure is calculated from the equation 
\begin{equation}
\label{eq:Kelvin}
    p = \frac{T {\rho}}{\mathcal{K} \mu}, \quad \mathcal{K}= \frac{m_u u_0^2}{k_B},
\end{equation}
where $k_B$ is the Boltzmann constant, $m_u$ is the atomic mass unit, $u_0$ is the initial outflow speed, and $\mu$ is the mean molecular weight. The conversion factor $\mathcal{K}$ then depends on the velocity, $u_0$, \citep{Mignone:14}.

We assume that the Cepheid wind is pulsation-driven, so we set the initial velocity to a slow outflow parametrized with the speed of sound at $t=0$ and then we drive the outflow though the pulsation velocity wave. The sound speed is dervied from the density and pressure profile, $c_s = \sqrt{\gamma \cdot p/\rho}$, and we set the initial outflow speed to be a slow subsonic value of 10\% of the speed of sound, $u_0 = 0.1 c_s$.

\paragraph{{\bf Boundary Conditions and Pulsation:}} All the simulations here are 1D and we only need to define two boundary conditions for our computational domain [1,500] $R_{\delta-Cep}$. 
The density at the inner boundary at $r = R_{\delta-Cep}$ is set to be equal to $\rho_0$, while the pressure is defined again from Eq.~\ref{eq:Kelvin}.
We have run a collection of simulations with different prescriptions for the pulsation wave. Our simulations can be divided into three categories and the pulsation is prescribed as either (a) a simple sinusoidal temporal function $u_p = u_0+u_{1}\cdot \sin{(\omega_{p} t)}$, or (b) the combination of two sinusoidal waves with different periods $= u_p = u_0+u_{1}\cdot \sin{(\omega_{p} t)} + u_{2}\cdot \sin{(2\omega_{p} t)}$, or (c) the first five terms of the Fourier series decomposition of the original radial velocity profile using only sinusoidal waves. 
The analytic equation of the decomposition of the observed radial velocity into a series of sinousoids after keeping the first 5 terms is:
\begin{eqnarray*}
     u(x_1) &=& u_0 + u_1 \cdot \sin{(\omega t) } 
           + u_2 \cdot \sin{(2\omega t)} + u_3 \cdot \sin{(3\omega t)} \nonumber \\
           &+& u_4 \cdot \sin{(4\omega t)} 
            + u_5 \cdot \sin{(5\omega t)}
\end{eqnarray*}
The constant $u_0=-16.40$~km s$^{-1}$ is the average speed from the observed radial velocity curve corresponding to the relative speed of $\delta$~Cep with respect to the Earth, thus we need to subtract this from the Fourier decomposition, since this term is not part of the pulsation driver. Thus the velocity at the inner boundary corresponding to the observed radial velocity is
\begin{eqnarray}\label{eq:Fourier5}
u(x_1) &=&  -15.32 \cdot \sin{(\omega t) } 
           -6.10 \cdot \sin{(2\omega t)} - 3.05 \cdot \sin{(3\omega t)} \nonumber \\
           &-&1.36 \cdot \sin{(4\omega t) }
           -0.49 \cdot \sin{(5\omega t)} \qquad [\mathrm{km\ s^{-1}}]
\end{eqnarray}

More specifically, our method had three main steps, namely
(a) Initially we used the astropy library to fit the observed RV profile. The astropy fourier decomposition algorithm has been tested on RR-Lyrae stars\footnote{The astropy routine can be found at: \url{https://docs.astropy.org/en/stable/timeseries/lombscargle.html}}.
(b) We used the fitted model to interpolate (upsample) the observed data.
(c) We built a simple algorithm that performs linear regression using the increasing order sinusoidal terms as independent features. With the help of this algorithm we expanded the original RV signal into the first 5 terms of a sinusoid series and tested that the decomposition into harmonics reproduced the original data within a satisfactory accuracy.
In section \ref{sec:results}, we describe each simulation separately.


\paragraph{{\bf Cooling:}} For the cooling function, \emph{PLUTO} allows for several different radiative cooling considerations, including for a power law and a tabular version. Here we use an analytic function (see, red curve in Fig.~\ref{fig:lambda}), which fits the cooling table, the values of which are plotted in blue curve in Fig.~\ref{fig:lambda} \citep[see][]{Tesileanu.etal:08}. The cooling table in \emph{PLUTO} is fitted by a polynomial of degree 10, $\Tilde{\Lambda}_{fit} = p_0 + p_1 x + p_2 x^2 + p_3 x^3 + p_4 x^4 + p_5 x^5 + p_6 x^6 + p_7 x^7 + p_8 x^8 + p_9 x^9 + p_{10} x^{10}$, as shown in the Fig.~\ref{fig:lambda}.

The characteristic cooling time for T$=10^7$~K and n=10$^9$~cm$^{-3}$ is:
\begin{equation}
   \tau_c =\frac{k_B T}{\Tilde{\Lambda}(T) n} \approx \frac{10^{-16} 10^{7}}{10^{-23} 10^{9}} sec= 1.3\times10^{5} sec,
\end{equation}
or 0.45 code units.
For n=10$^{6}$~cm$^{-3}$, the cooling time becomes $\tau_c=1.3\times10^{8}$~s or 462 code units, i.e. pulsation periods. The first option would be the only one to work in our case, in other words we want the cooling to take place in a timescale equivalent to a fraction of the pulsation period.

The characteristic cooling time for T=$5\times10^4$~K and n=10$^9$~cm$^{-3}$ is $\tau_c=29$~s or 0.0001 code units.
For n=10$^{6}$~cm$^{-3}$, the cooling time becomes $\tau_c=2.9\times10^{4}$~s or 0.1 code units. The last option would be the only one to work in our case, in other words we want the cooling to take place in a timescale equivalent to a fraction of the pulsation period that corresponds to a time interval of the order of one day, to reproduce the observed X-ray increase and decay. A more detailed discussion on characteristic time-scales can be found in the Discussion section~\ref{sec:discussion}.


\subsection{Results}\label{sec:results}
In Table~\ref{tab:sims} we present all the numerical tests and experiments performed so far. 
A steady state stellar wind type outflow is not a necessary first step, as the pulsation will dominate and drive the outflow giving the same picture as shown in \citet{Boulangier.etal:19}. 
For that reason, we do not have to use any additional heating terms in our energy equation to heat the stellar corona and accelerate the stellar wind.
For all our simulations, we start from the predefined initial and boundary conditions and let each simulation reach a steady state after 1000 pulsation periods. Once we reach the steady state, we then zoom into individual pulsation periods and examine the HD solution, the mass-loss rate and the synthetic observations corresponding to each solution.
{Below we describe the most important simulation results. A detailed description of each separate simulation run of Table~\ref{tab:sims} can be found in the Appendix \ref{sec:appendix}.}

\begin{deluxetable*}{lcccc|cccc}[htpb]
\tablecaption{Different simulations of the Cepheid atmosphere using hydrodynamic one-dimensional setups in spherical geometry, with $\rho_0=10^9 cm^{-3}$, $L_0 = 1 erg\ s^{-1}\ cm^{-2}$, $A = 3.0\cdot c_s$. \label{tab:sims}}
\tablecolumns{9}
\tablenum{1}
\tablewidth{0pt}
\tablehead{
\colhead{Run \#} & \colhead{$u(x_1)$ [100 km s$^{-1}$]} & \colhead{$T_0$ [K]} & Cooling? & \colhead{$\rho(x_1)$ [$\rho_0$]} & \colhead{$u(\infty)$  [km s$^{-1}$]}  & \colhead{$\dot{M}\ [M_\odot\ yr^{-1}]$} & \colhead{$L_{min}\ [L_0]$} & \colhead{$L_{max}\ [L_0]$}}
\startdata
01 & 10 $\times$ Eq.~\ref{eq:Fourier5} & $5\times10^5$ &  No & $10^{-3}$ & 80 & $1.36\times 10^{-11}$ & $8.95\times10^{-18}$ & $8.39\times10^{-16}$ \\
02 & 10 $\times$ Eq.~\ref{eq:Fourier5} & $5\times10^5$ &  Poly - Fit & $10^{-3}$ & 40 & $1.62\times 10^{-13}$ & $1.09\times10^{-20}$ & $1.17\times10^{-16}$ \\
03 & 10 $\times$ Eq.~\ref{eq:Fourier5} & $5\times10^5$ & Tabular &  $10^{-3}$ & 60 & $2.49\times 10^{-13}$ & $2.32\times10^{-21}$ & $1.73\times10^{-16}$ \\
04 & $A \cdot \sin{(\omega t)}$ & $5\times10^5$ & Tabular &  $10^{-3}$ & 200 & $2.65\times 10^{-11}$ & $5.35\times10^{-20}$ & $4.53\times10^{-16}$\\
05  & $A \cdot \sin{(2 \omega t)}$ & $5\times10^5$ & Tabular & $10^{-3}$ & 200 & $2.65\times 10^{-11}$ & $4.48
\times10^{-19}$ & $3.67\times10^{-16}$ \\
06 & $A \cdot \sin{(\omega t)}$ & $2\times10^6$ & Tabular & $10^{-3}$ & 580 & $5.55\times 10^{-11}$ & $8.83\times10^{-19}$ & $1.60\times10^{-15}$ \\ 
07 & $A\cdot \sin{(\omega t)} + A\cdot \sin{(2 \omega t)}$ & $5\times10^5$ & Tabular & $10^{-3}$ & 400 & $3.30 \times 10^{-11}$ & $3.12\times10^{-19}$ & $1.11\times10^{-15}$ \\
08 & $A/2\cdot \sin{(\omega t)} + A/2\cdot \sin{(2\omega t)}$ & $5\times10^5$ & Tabular & $10^{-3}$ & 110 & $1.79 \times 10^{-11}$ & $2.46\times10^{-19}$ & $5.97\times10^{-16}$ \\
09 & $A\cdot \sin{(\omega t)} + A/2\cdot \sin{(2\omega t)}$ & $5\times10^5$ & Tabular & $10^{-3}$ & 280 & $2.74 \times 10^{-11}$ & $3.84\times10^{-19}$ & $1.13\times10^{-15}$ \\
10 & $A\cdot \sin{(\omega t)} + A/4\cdot \sin{(2\omega t)}$ & $5\times10^5$ & Tabular & $10^{-3}$ & 230 & $2.65 \times 10^{-11}$ & $2.14\times10^{-19}$ & $6.66\times10^{-16}$ \\
11 & $A \cdot \sin{(\omega t)}$ & $2\times10^6$ & Tabular &  $10^{-2}$ & 560 & $6.10 \times 10^{-10}$ & $5.03\times10^{-18}$ & $1.34\times10^{-13}$ \\
12 & $A \cdot \sin{(\omega t)}$ & $5\times10^5$ & Tabular &  $10^{-2}$ & 170  & $2.67 \times 10^{-10}$ & $1.26\times10^{-25}$ & $4.99\times10^{-17}$  \\
13 & $A \cdot \sin{(0.5\omega t)}$ & $5\times10^5$ & Tabular &  $10^{-3}$ & 200  & $2.61 \times 10^{-11}$  &  $1.95\times10^{-20}$ & $3.19\times10^{-16}$  \\
14 & $A \cdot \sin{(0.25\omega t)}$ & $5\times10^5$ & Tabular &  $10^{-3}$ &  200  & $1.63 \times 10^{-11}$  & $1.17\times10^{-20}$  & $2.13\times10^{-16}$  \\
$15^{\star}$ & $A \cdot \sin{(\omega t)}$ & $5\times10^5$ & Tabular &  $10^{-1}$ &  160  & $3.72 \times 10^{-10}$ & -  & - \\
16 & $A \cdot \sin{(\omega t)}$ & $5\times10^5$ & No &  1 &  230  & $2.64 \times 10^{-8}$ & $1.41\times10^{-14}$ & $1.39\times10^{-12}$  \\
17 & $A \cdot \sin{(\omega t)}$ & $5\times10^5$ & No &  $10^{-1}$ & 230  & $2.64 \times 10^{-9}$ & $2.98\times10^{-15}$ & $7.57\times10^{-13}$  \\
18 & $A \cdot \sin{(\omega t)}$ & $5\times10^5$ & No &  $10^{-2}$ & 230 & $2.64 \times 10^{-10}$ & $4.97\times10^{-17}$ & $4.90\times10^{-14}$  \\
19 & $A \cdot \sin{(\omega t)}$ & $5\times10^5$ & No &  $10^{-3}$ &  230  & $2.64 \times 10^{-11}$ & $5.25\times10^{-19}$ & $6.33\times10^{-16}$  \\
20 & $A \cdot \sin{(\omega t)}$ & $5\times10^5$ & Tabular &  $10^{-3}\cdot (R_\star/r)^{4}$ & 200 & $2.65\times 10^{-11}$ & $5.35\times10^{-20}$ & $4.53\times10^{-16}$\\
\enddata
\tablecomments{$^{\star}$Run 15 was relaxed only for 550 pulsation cycles, enough for the initial perturbations to reach the outer boundary. After that the simulation crashed due to very low temperatures close to the star. \clearpage}
\end{deluxetable*}


\paragraph{{\bf Run 01}} 

\begin{figure*}[htbp]
 \begin{center}
\includegraphics[width=\textwidth,trim={0cm 1cm 0cm 0cm},clip]{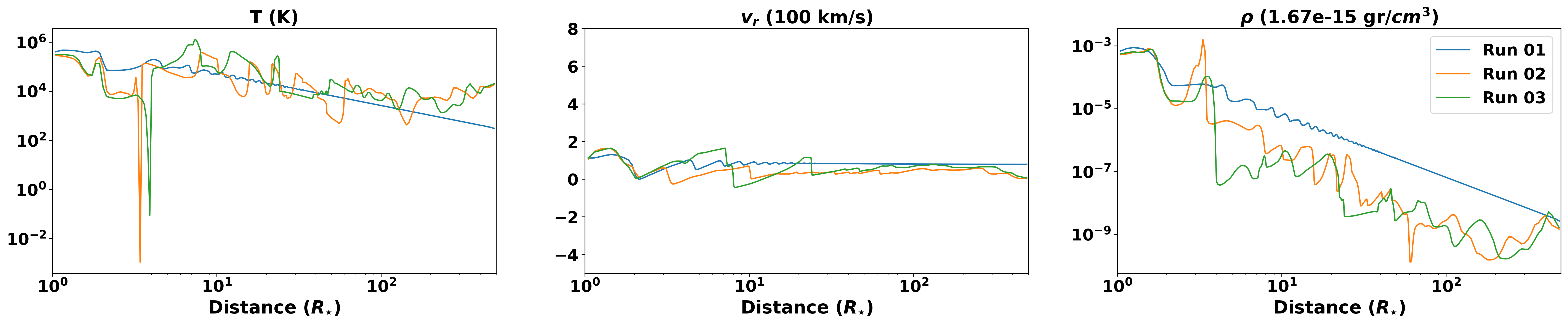}
\includegraphics[width=\textwidth,trim={0cm 1cm 0cm 0cm},clip]{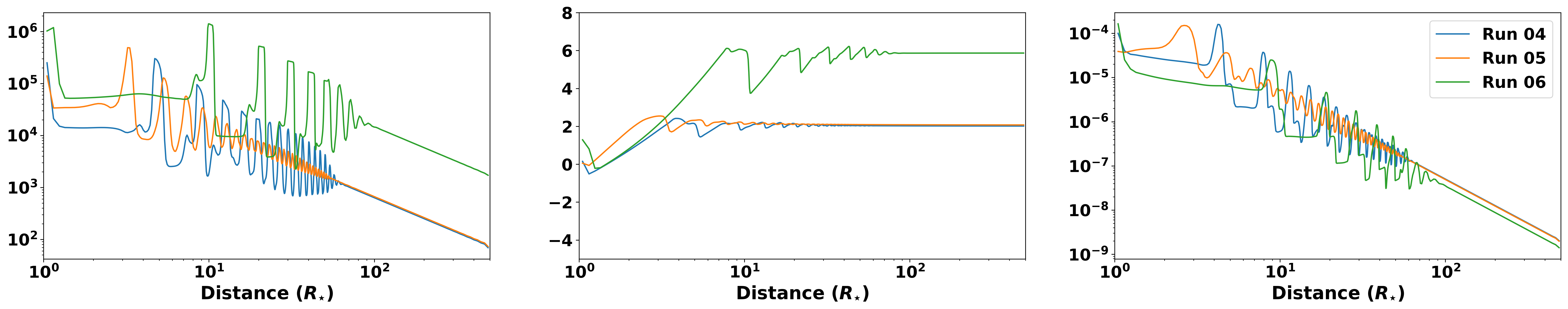}
\includegraphics[width=\textwidth,trim={0cm 1cm 0cm 0cm},clip]{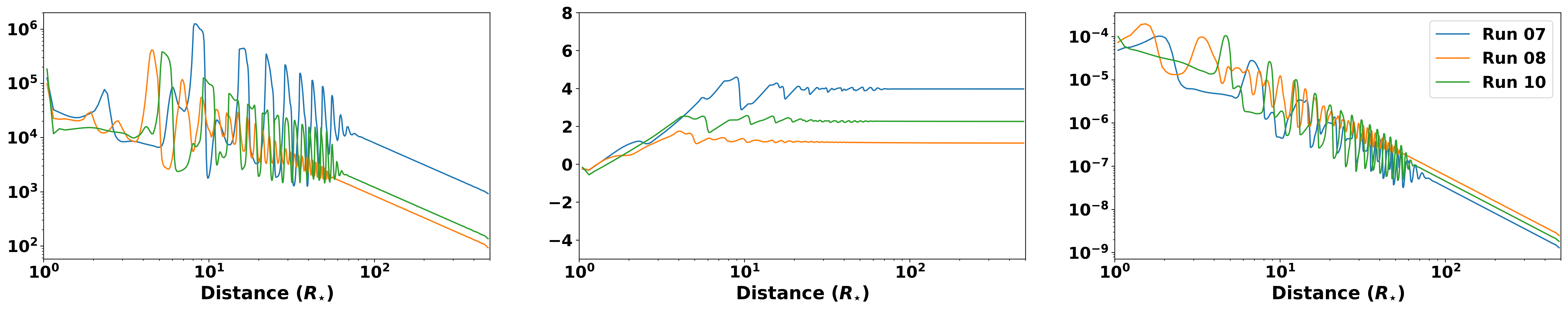}
\includegraphics[width=\textwidth,trim={0cm 1cm 0cm 0cm},clip]{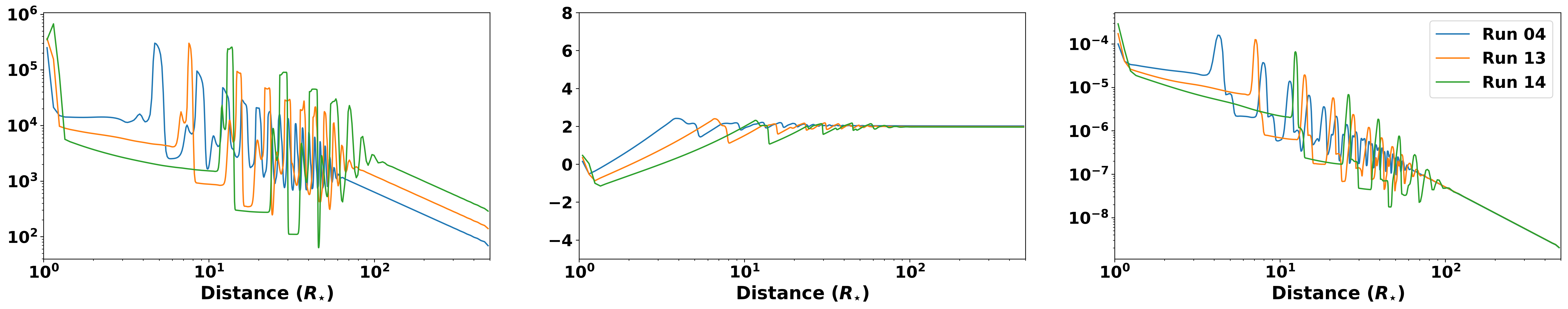}
\includegraphics[width=\textwidth,trim={0cm 1cm 0cm 0cm},clip]{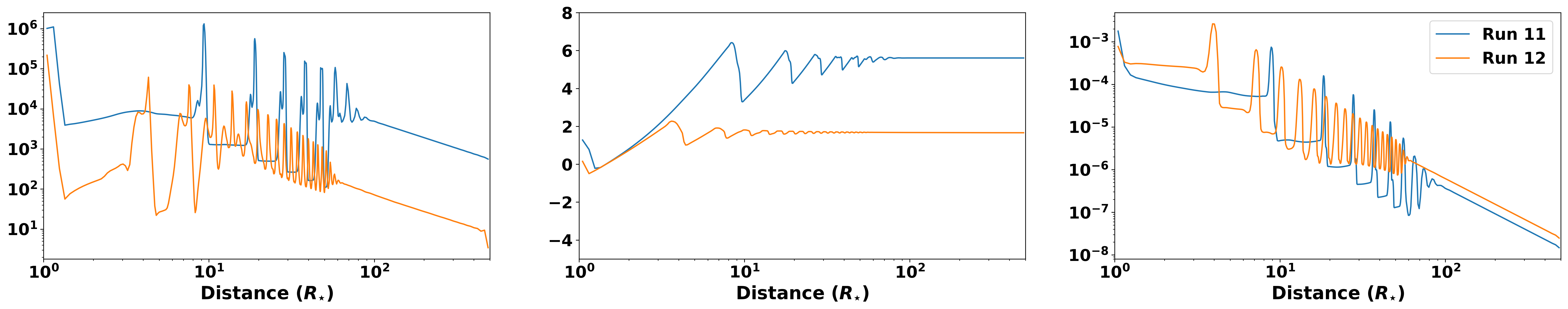}
\includegraphics[width=\textwidth,trim={0cm 0cm 0cm 0cm},clip]{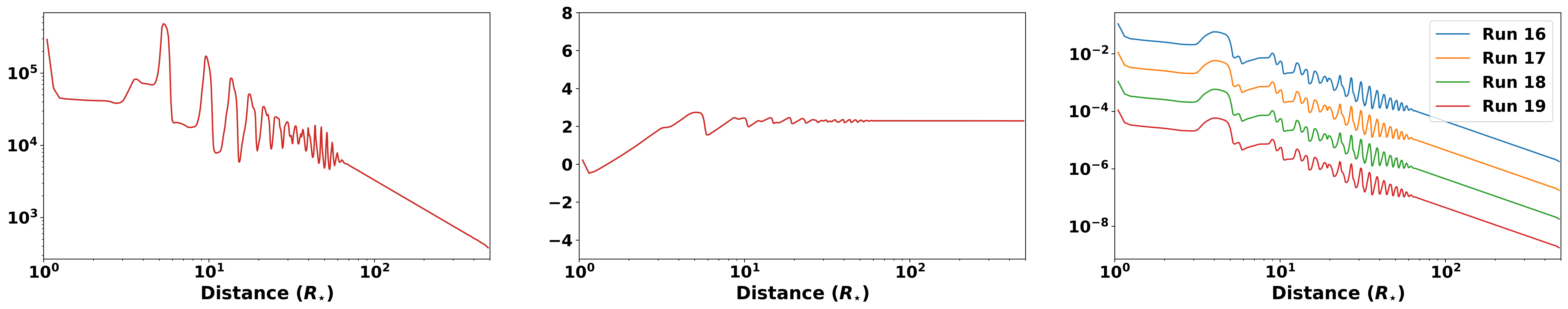}
  \caption{Run 01 - 19: Converged solutions in spherical coordinates, with gravity after 1000 pulsation cycles showing the radial dependence of temperature (left column), velocity (middle) and density (right).}
  \label{fig:1000Puls}
  \end{center}
  \end{figure*}

In Fig.~\ref{fig:1000Puls}, we show the initial and converged simulation state after 1000 pulsation cycles in spherical geometry including the gravitational attraction of  $\delta$~Cep in the ideal case, i.e. \textit{without cooling}. We wanted to examine whether the actual observed radial velocity profile would significantly affect the hydrodynamic solution of the Cepheid atmosphere. For that reason we expanded the original radial velocity profile using the sinusoidal signals with periods similar to the Fourier expansion given by Eq.~\ref{eq:Fourier5}. 

The resulting steady state solution is shown in the top panel of Fig.~\ref{fig:1000Puls}. The temperature in the left panel is decreasing outwards as a power law, similar to the density profile shown in the right panel. More specifically, the steady state convergent solution creates a density profile that drops as $\rho\propto r^{-2}$ in the gravitational field of the Cepheid. 
The temperature profile is then expected to be $T\propto P/\rho=\rho^{\gamma}/\rho=\rho^{\gamma-1}=r^{-4/3}$. And this is indeed the power law that the steady solution follows. The velocity shows some initial oscillations due to the pulsation driver at the inner boundary, which gradually decay, and a terminal outgoing speed of about 80 km s$^{-1}$ is reached. 

\begin{figure*}[htbp]
 \begin{center}
   \includegraphics[width=0.45\textwidth,trim={0cm 0cm 0cm 0cm},clip]{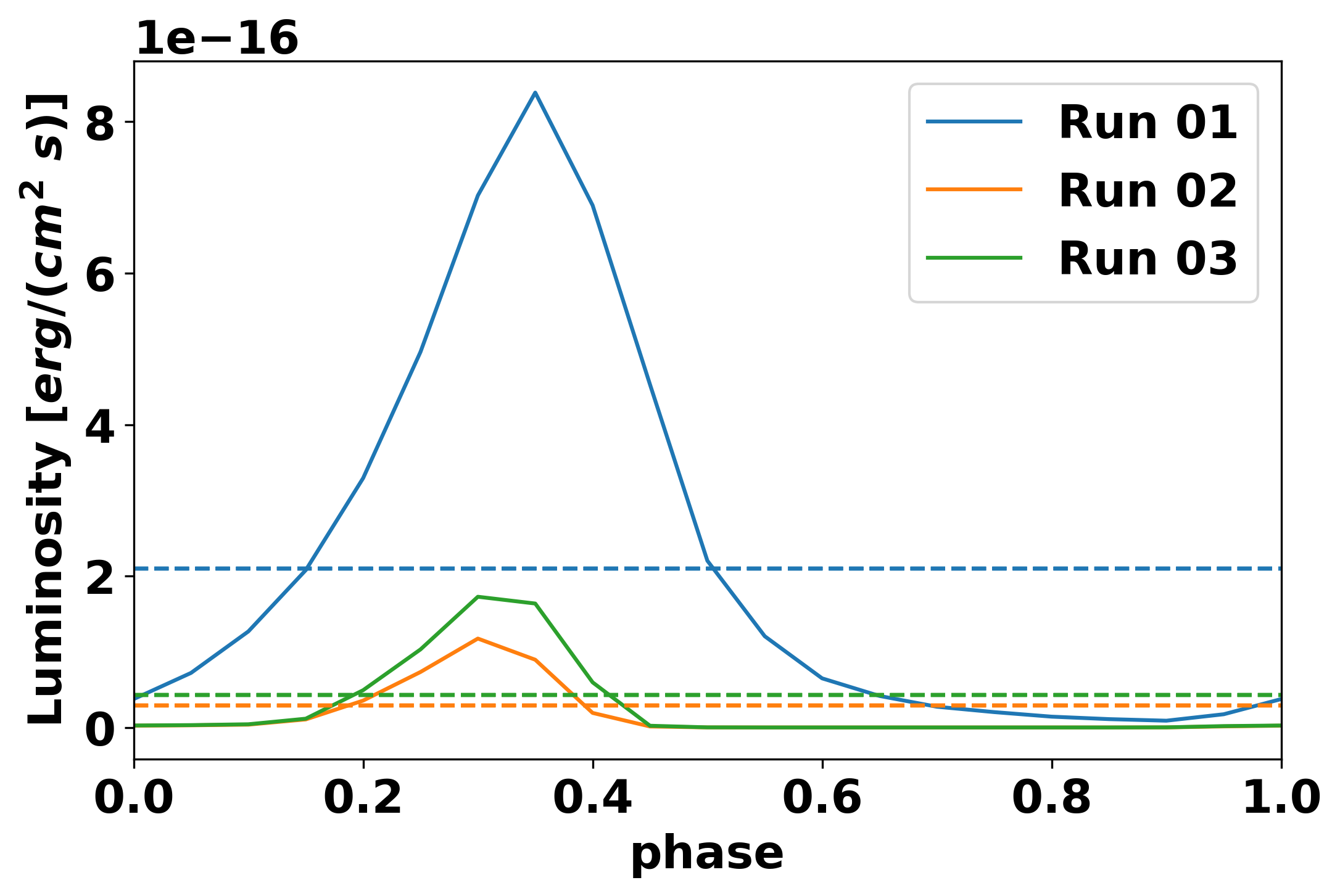}
    \includegraphics[width=0.45\textwidth,trim={0cm 0cm 0cm 0cm},clip]{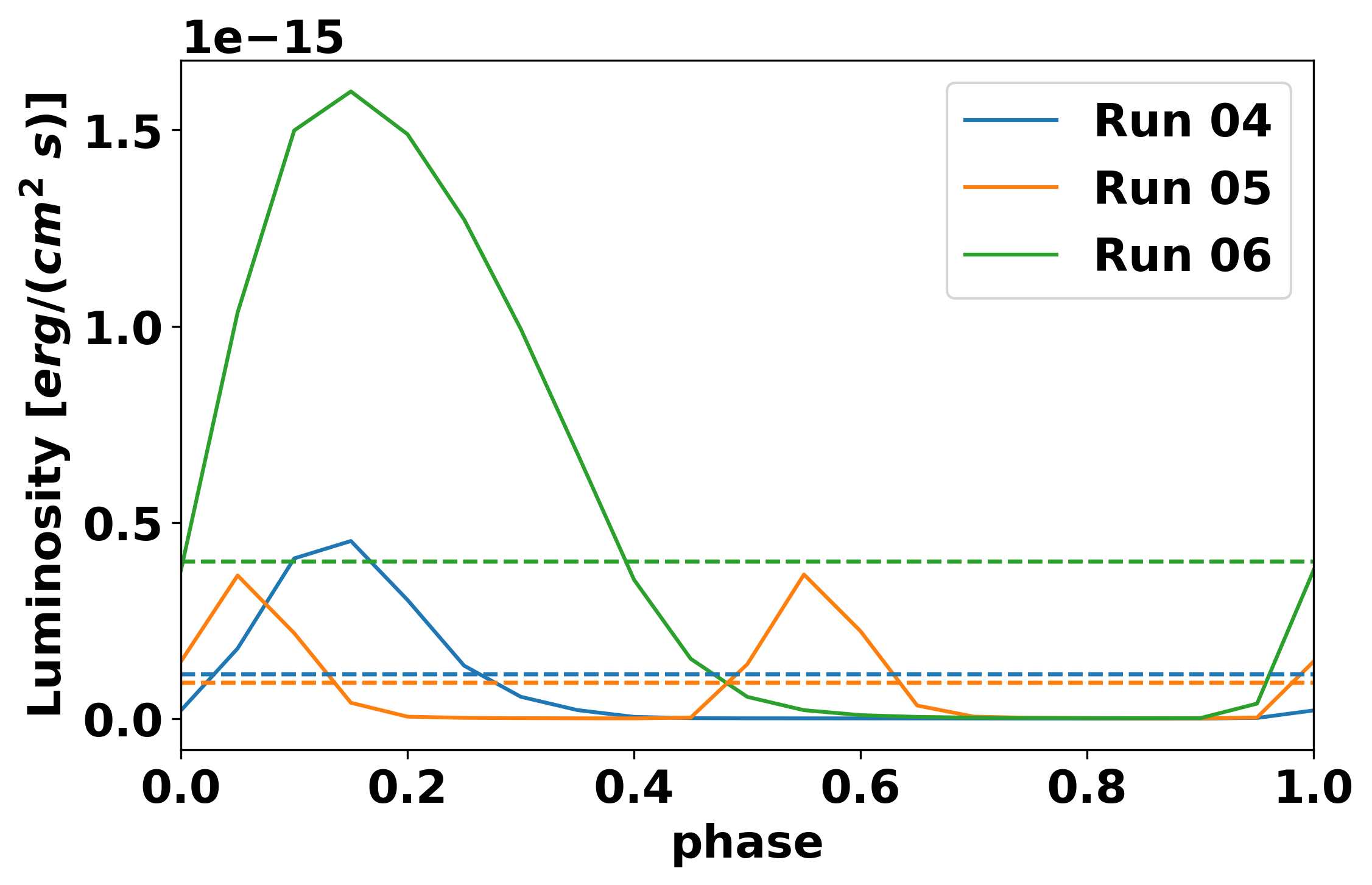}
    \includegraphics[width=0.45\textwidth,trim={0cm 0cm 0cm 0cm},clip]{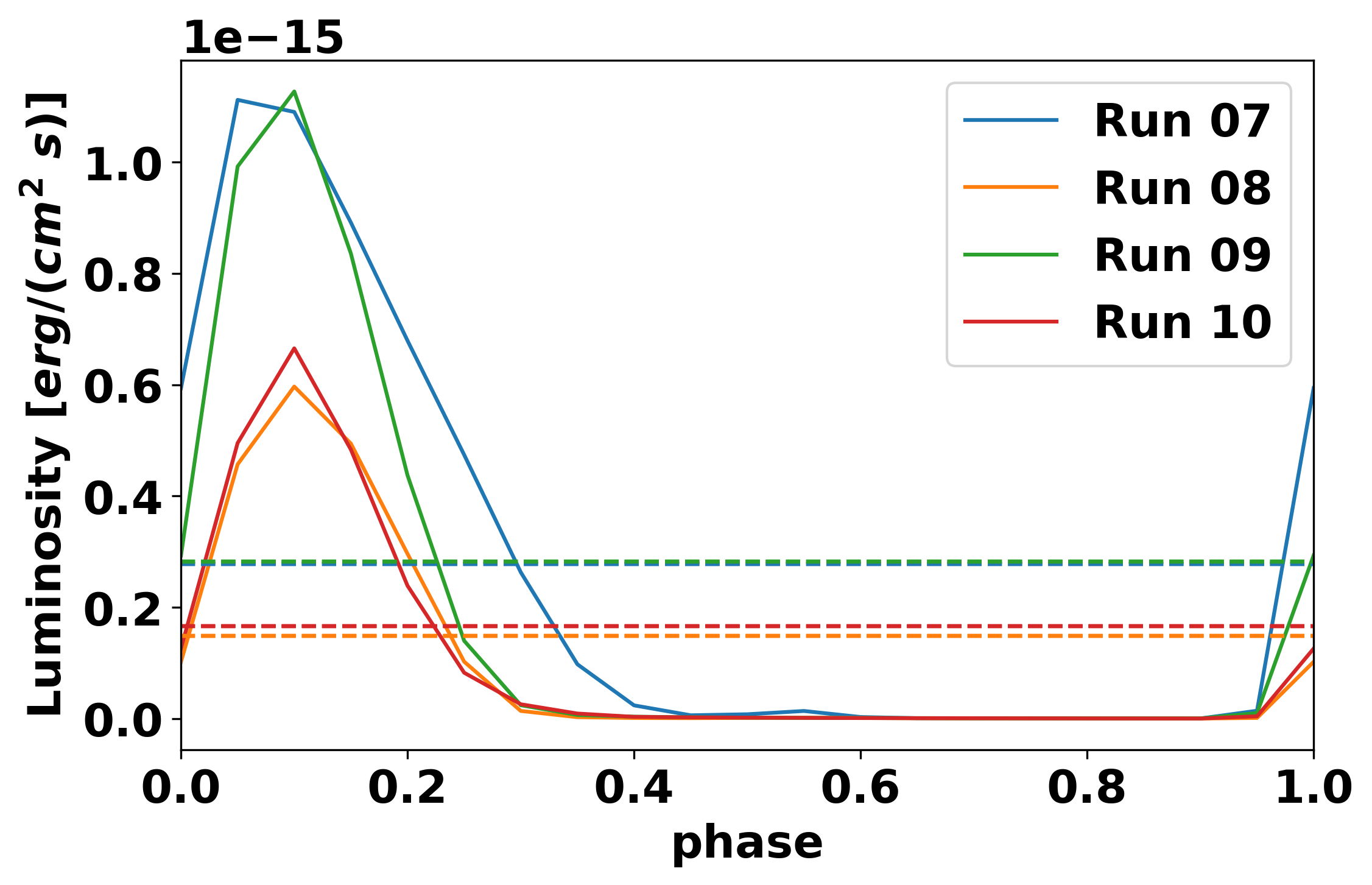}
    \includegraphics[width=0.45\textwidth,trim={0cm 0cm 0cm 0cm},clip]{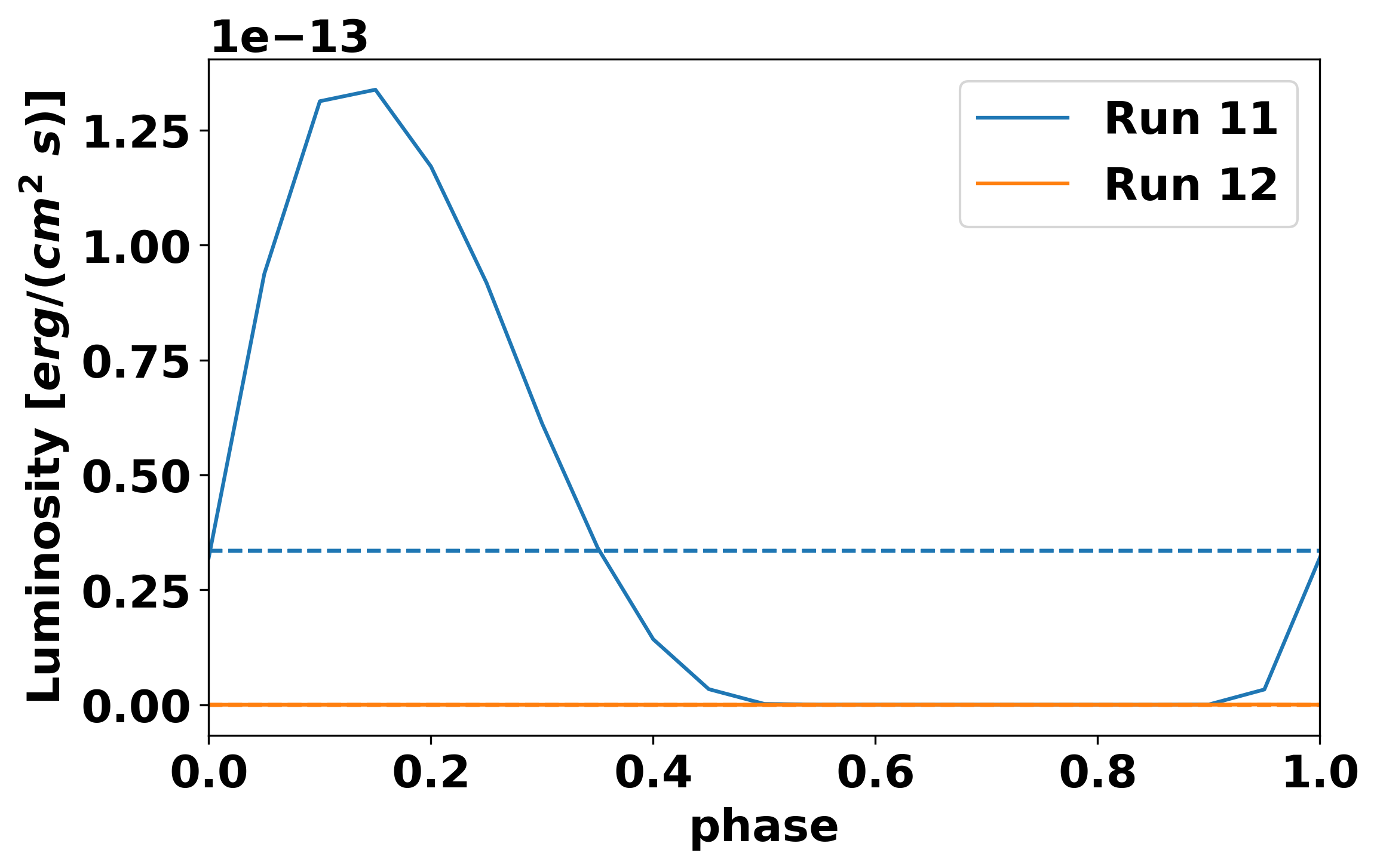}
    \includegraphics[width=0.45\textwidth,trim={0cm 0cm 0cm 0cm},clip]{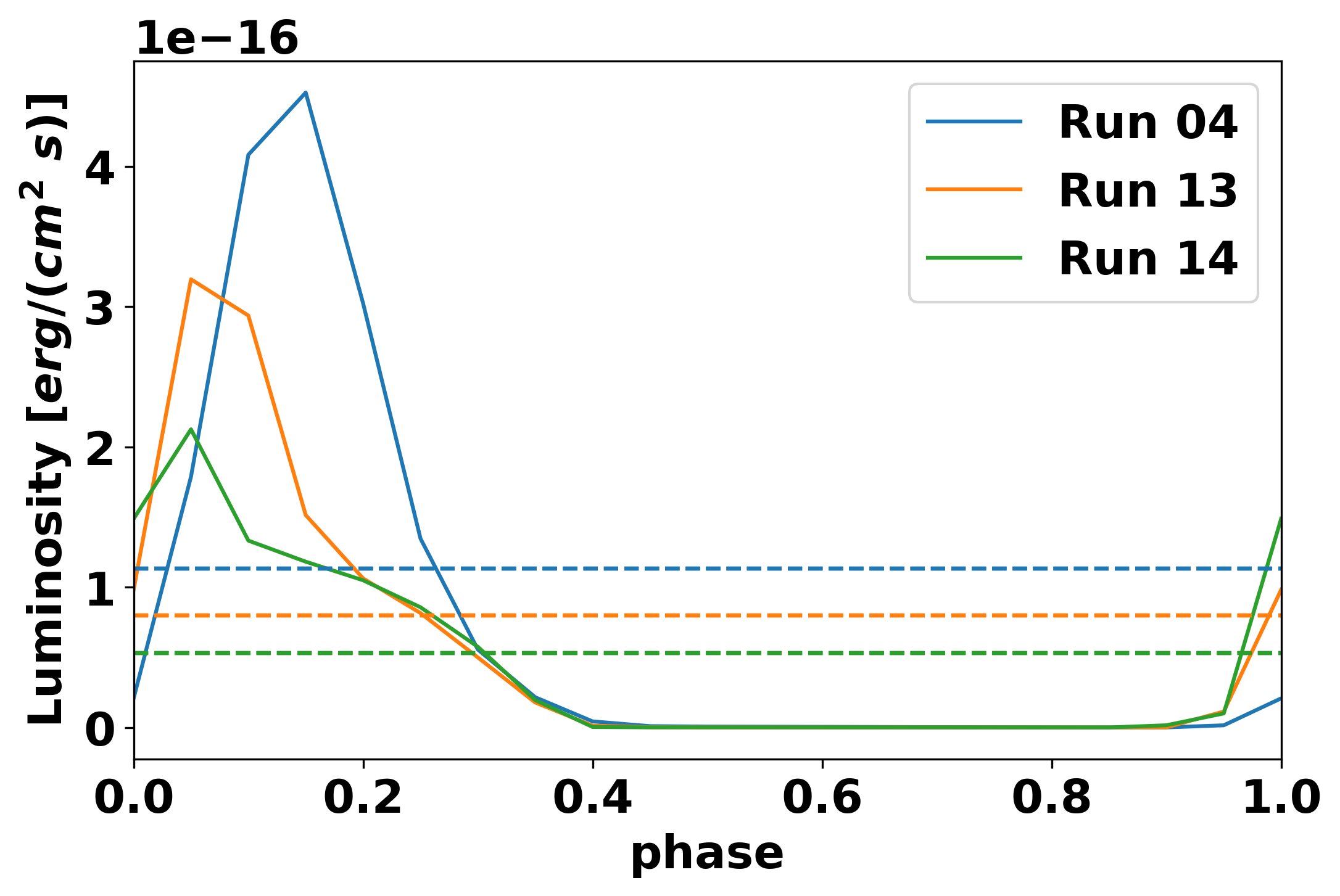}
    \includegraphics[width=0.45\textwidth,trim={0cm 0cm 0cm 0cm},clip]{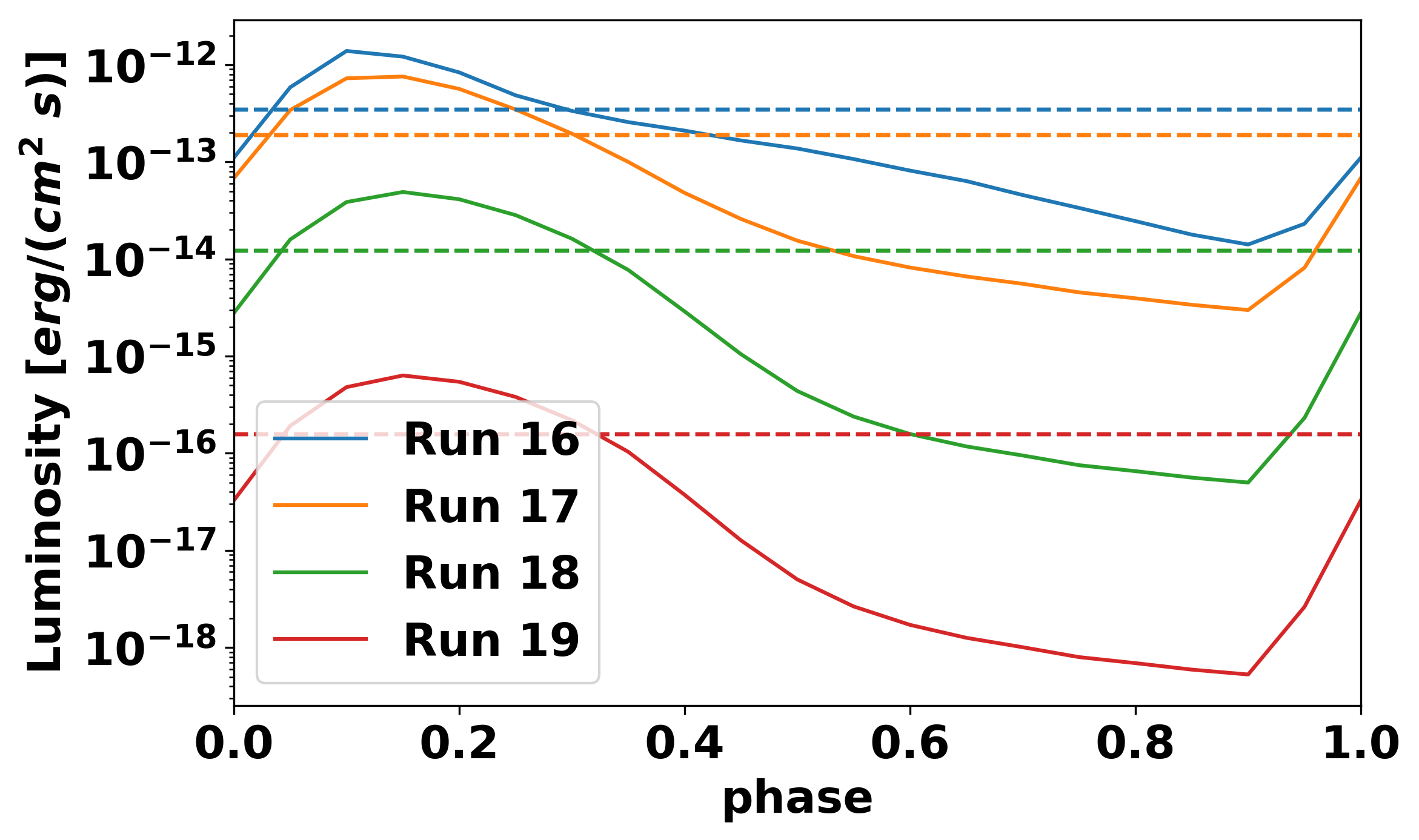}
  \caption{Run 01 - 19: X-ray light curves during one pulsation cycle. Phases have been adjusted to agree with the observationally defined phase based on maximum optical luminosity. }
  \label{fig:lightcurve}
  \end{center}
  \end{figure*}

 In Fig.~\ref{fig:lightcurve}, we have adjusted the phase of the results to agree with the observational radial velocity curve corresponding to maximum light. The Fourier cases at simulation t=0 are at zero radial velocity moving towards positive values for the observer, i.e. inflow for the star, (0.4 phase difference with observations), while all other cases start from zero radial velocity moving towards negative values for the observer, i.e. outflow, (0.9 phase difference with the observations). The luminosity is calculated and normalized at the distance of $\delta$~Cep and all plots have been shifted with $\phi=0$ corresponding to maximum visible light. {In Fig.~\ref{fig:lightcurve}, we also indicate with dashed horizontal lines where the luminosity level of quarter maximum lies, i.e. 4 times lower than the peak luminosity, for each simulation in the same corresponding color as the full light curve in order to better compare with the observations \citep{Engle.etal:14}. The light curve obtained shows one peak around phase 0.3-0.4, with the luminosity varying by about 2 orders of magnitude.}

    In Fig.~\ref{fig:1000Puls} we show a simulation after 1000 pulsation cycles with a base temperature T$=5\times10^5$~K and density of $n_e=10^6$~cm$^{-3}$. We see sharp peaks of the temperature appearing and reaching its peak temperature close the the star, and reaching several MK temperatures matching the inferred plasma temperature range from X-rays. At the same time the density bumps decay fast, as expected and similar to the case without cooling.

\paragraph{{\bf Runs 02 - 03}} 
{For these simulations, we use similar setups as for Run 01, with the difference that for Run 02 we include a radiative cooling function given by the polynomial fit presented in section \ref{sec:methods} and for Run 03  the cooling is given by a tabulated form as discussed in \citet{Mignone:14} and shown by the blue curve in Fig.~\ref{fig:lambda}.
In the simulations with cooling included, we need an even stronger shock to keep seeing the same bumps in the temperature and density profiles, as for Run 01, when cooling was excluded. 
In this case, the profiles of temperature, and density are much more disturbed, with major cooling taking place around $r=2 R_\star$ and the temperature dropping dramatically.
The polynomial fit follows closely the more accurate tabulated form of cooling as shown in Fig.~\ref{fig:lambda}.
The tabulated cooling function is more accurate with a similar computational cost as the polynomial function, so we decided to use the tabulated cooling for all the cases hereafter.}

\paragraph{{\bf Run 04}} The same setup as Run 03 is used, with the exception that for the velocity we now have a simple sinusoidal function with amplitude $A = 3 c_s$ and a frequency corresponding to the pulsation frequency $\omega = 2\pi/T_p$. This parametric form of the outflow ensures a large enough speed that can escape the gravitational attraction of the star. For this case, the local sound speed at $r=R_\star$ is about 100~km s$^{-1}$ and the pulsation amplitude is set to 3 times that, i.e. 300 km s$^{-1}$, which is larger than the local escape speed and thus the material escapes and a steady state outflowing wind solution is formed. The terminal speed of the converged solution is 200~km~s$^{-1}$, similar to the shock speeds suggested by \citet{Engle.etal:17}.

This solution is significantly different compared to the Fourier decomposition cases, in that the driving velocity perturbation is now larger. The perturbation enforced at the inner boundary is then creating characteristic oscillations in the profiles of all the quantities that gradually decay at a distance of about 100~$R_\star$.

    \begin{figure*}[htbp]
 \begin{center}
   \includegraphics[width=\textwidth,trim={0cm 0cm 0cm 0cm},clip]{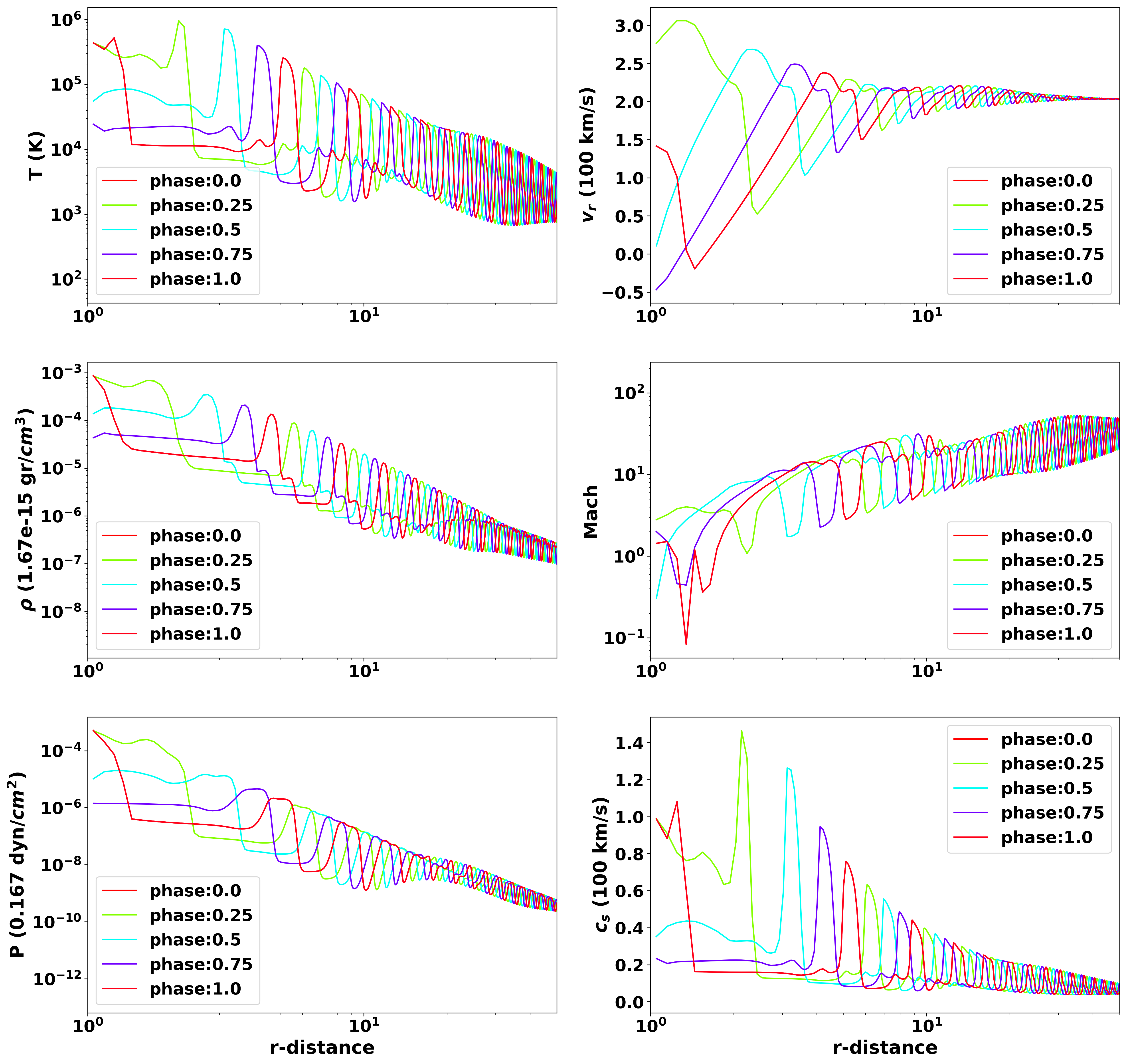}
  \caption{Run 04: Zooming in on one period for all HD quantities.}
  \label{fig:phasecool}
  \end{center}
  \end{figure*}


\paragraph{{\bf Run 06}} For this run, we use the same setup as Run 04, but with a background temperature of $T=2\times10^6$K. This leads to a distinctly different solution. The terminal speed is now significantly different, 600 km s$^{-1}$, i.e. a factor 3 higher, compared to Run 04. The first peaks in all profiles appear later on, as for a warmer plasma the characteristic sound speed is larger and the perturbations propagate faster. A higher average temperature and density are reached at larger distance, as shown in the second row of Fig.~\ref{fig:1000Puls}. The peak of the light curve for Run 06 is visibly higher than Run 04 and Run 05, as shown at the top right panel of Fig.~\ref{fig:lightcurve}.

\paragraph{{\bf Run 07}} We use the same setup as for Run 04, but this time instead of a simple sinusoidal function for the velocity, we use the combination of two sinusoidal waves with the same amplitude, $A$, and two different frequencies $\omega$ and $2\omega$. The terminal speed reached is now double that compared to Run 04 and Run 05, i.e. 400 km s$^{-1}$, as the velocity perturbation amplitude has now doubled.





\paragraph{{\bf Run 12}} For this run, we use the same setup as Run 11, but for a coronal temperature of 500,000K. The terminal speed is 270 km~s$^{-1}$, the perturbation amplitudes are lower and the peaks appear earlier. The temperature at distances less than 10~$R_\star$ are very low, reaching a few tens of K. The corresponding peak of the light curve in Fig.~\ref{fig:lightcurve} is significantly lower compared to the hot case of Run 11.

\paragraph{{\bf Runs 16 - 19}} Finally, we switched the cooling off and ran 4 cases based on Run 04 with the same density, 10 times, 100 times, and 1000 times larger base densities than that of Run 04, as shown in Table~\ref{tab:sims}. The results are very interesting. All terminal speeds are now 230 km s$^{-1}$, and the computed mass-loss rates are differing by exactly one order of magnitude between each case progressively. A similar trend is also followed in the X-ray light curves. Fig.~\ref{fig:1000Puls} shows the reason for this. The density and thus the pressure profiles are exactly the same, just differing by an order of magnitude, thus the temperature and the sound speed, that depend on the ratio $P/\rho$, are identical for all four cases. Subsequently, the terminal speed which depends on the initial perturbation and the sound speed are also identical.


   In order to conclude whether the density and temperature bumps are phase dependent and thus appropriate to explain the X-ray  observations \citep{Engle.etal:17}, we zoom in on one of the simulated pulsation cycles as shown in Fig.~\ref{fig:phasecool}. The temperature appears to go from very low, around a few tens of thousands Kelvin, to 20 MK during a single pulsation period. {This is evidence that the shock wave mechanism can produce the characteristic modulation in the coronal plasma of a Cepheid that causes the observed emission signature.} 

\subsection{Synthetic Light Curves and mass-loss}

The final step is to calculate in post-processing the synthetic X-ray light curves from our simulations following the same approach as in \citet{Pinto.etal:15}. More specifically, once the spatial distributions of density and temperature are known we can easily compute the emission measure of each computational cell, $EM(T)=n_k^2\Delta V_k$, with $n_k$ the plasma density and $\Delta V_k$ the volume of each cell. Then, the thermal X-ray emission of a plasma with known number density $n$ and temperature T at a given photon energy h$\nu$ can be computed \citep[see, e.g.][]{Pinto.etal:15} using an optically-thin, collision-dominated radiative loss model, such as CHIANTI or APEC \citep{Dere.etal:97,Dere.etal:19}.

     \begin{figure*}[htbp]
 \begin{center}
   \includegraphics[width=0.9\textwidth,trim={0cm 0cm 0cm 0cm},clip]{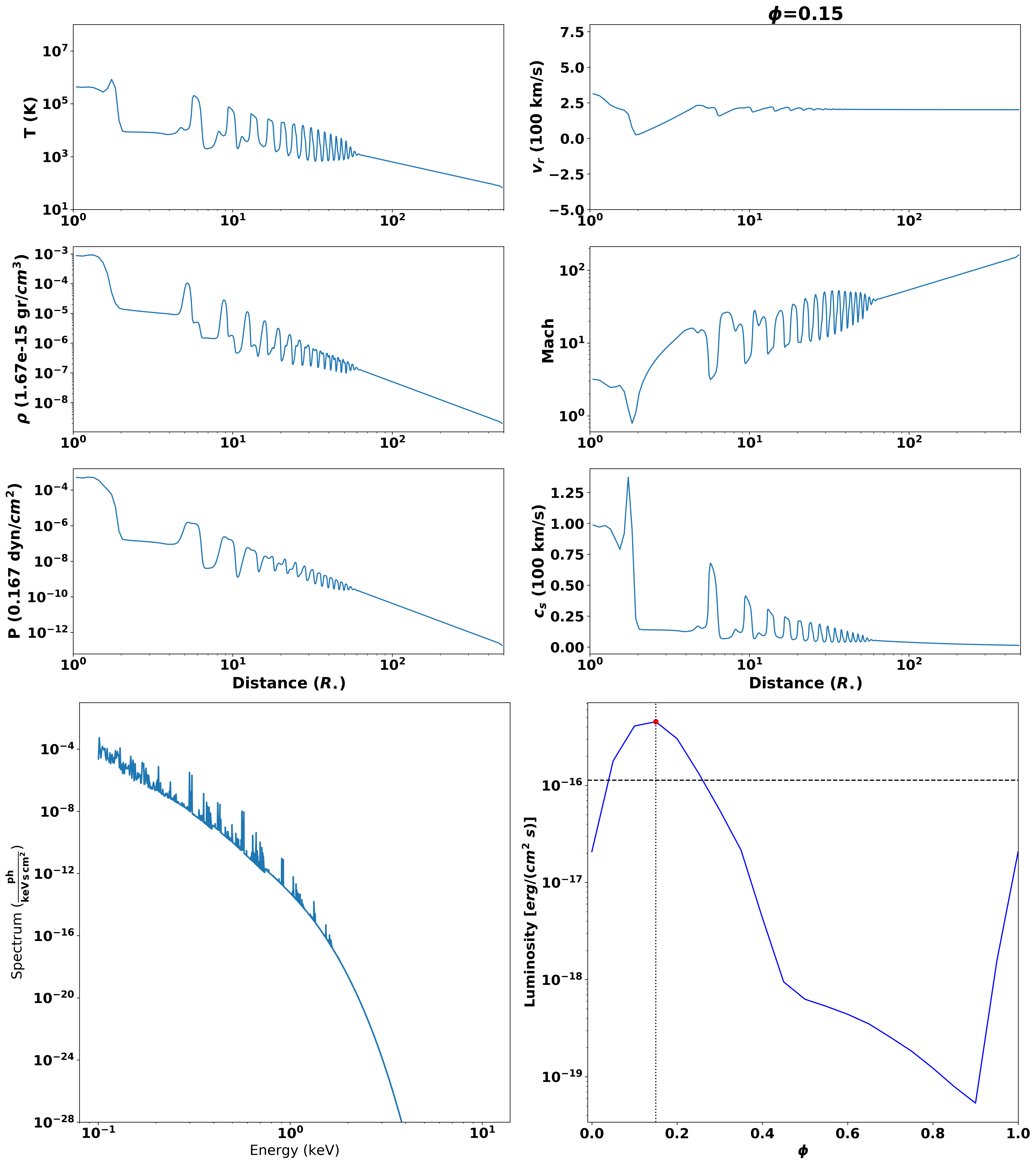}
  \caption{Run 04: \textit{Top panels}: Hydrodynamic simulation results. \textit{Bottom left}: Total spectra per simulation snapshot/pulsation phase. \textit{Bottom right}: X-ray light-curve for a full pulsation cycle. {An animation of the HD quantities, the total spectra, and X-ray light curve during one full pulsation cycle is available in HTML form in the online version of the journal. The time-step between each animation frame corresponds to a time interval of 0.05 $\phi$, i.e. there are 20 snapshots covering a full pulsation cycle. The phase of each frame is shown at the top right corner of each frame. The static version of the figure corresponds to phase 0.15 $\phi$, during which the X-ray light curve (bottom right) reaches its peak emission value, as determined by HD quantities (top) and reflected by the total spectra (bottom left). The peak X-ray emission is achieved when the recently emerged shock wave is at radial distance $r=2R_\star$.}}
  \label{fig:spectra}
  \end{center}
  \end{figure*}

We used the mock X-ray observation package Simulated Observations of X-ray Sources to create synthetic spectra for each simulation (SOXS\footnote{Available at \url{http://hea-www.cfa.harvard.edu/~jzuhone/soxs/}}). We first calculated the emission measure during a pulsation cycle and then calculated the thermal emission of the plasma in our physical domain with SOXS, see e.g. Fig.\ref{fig:spectra}. We then integrated the spectrum over the domain and obtained the X-ray luminosity at each time step during the simulation together with the X-ray light curve. The X-ray light curves for each numerical experiment are shown in Fig.~\ref{fig:lightcurve}. The luminosity is calculated and normalized at the distance of $\delta$~Cep.

The results of the 1D model in spherical geometry can be generalised to a spherically symmetric flow in 3D. We computed the mass-loss rate at each point of the simulation domain,  $\dot{\mathrm{M}}=\rho u_\infty r^2$, and obtained the total mass-loss rate as the value of the mass-loss profile at some distance beyond 100 $R_{\delta-Cep}$, as this is the mass eventually escaping. The mass-loss rates for each run can be seen in Table~\ref{tab:sims}.

\section{Discussion}\label{sec:discussion}

Shock waves are ubiquitous in the universe as sources of X-ray emission, from interactions of clusters of galaxies to supernova remnants to shock-driven winds of massive stars. 
In this study, we ran a large set of simulations in order to explore whether pulsation-driven shocks that form close to the Cepheid surface can explain the observed X-ray variability in short period Cepheids, with $\delta$~Cep serving as a case study. We explore various aspects of these simulations and their results below.

\subsection{Model Appropriateness}

Our simulations employed a simplified pulsation driver and we essentially performed  parametric studies on coronal temperature, base density and driving velocity function.
There are a number of aspects of these models that warrant some further examination or justification.

\paragraph{The need for numerical models}
Firstly, one might question whether full numerical models are required at all.  Recently, \citet{Prudil.etal:20} analysed the effects of shocks on the optical light curves of fundamental-mode RR Lyrae stars and found they can produce characteristic ``humps'' and ``bumps'' resulting from multiple shocks forming and propagating close to the stellar photosphere. Evidence for colliding and merging shocks is reproduced by our simulations as well, in Runs 04 - 14 and the runs without cooling 16 - 19. This is an indication of shock interactions, with more recently formed shocks ``catching up'' with shocks that are in their propagation path, and in the interaction region there are local peaks forming that eventually merge with the older shocks. The shock dynamics are very complex and this is evidence that we cannot solve the problem analytically or with simpler models, since we are only able to capture realistic shock evolution and interaction in time by including the gravitational force accurately.

\paragraph{Pre-existing hot plasma}
The shock wave scenario explored  requires a preexisting hot stellar corona in order for the X-ray emission that is observed to be produced through a thermal mechanism and have the emitting plasma reach 5-20 MK. To justify this condition, we appeal to the fact that non-pulsating supergiants have a similar level of quiescent X-ray emission as Cepheids, as shown by Ayres (2017) and noted in Section~\ref{S:1}. It thus seems likely that the existing hot plasma prerequisite is indeed present.


\paragraph{Pulsation velocity amplitude}
In order to produce a pulsation driven outflow without significant inflows we used a pulsation amplitude that is parametrised with the local sound speed as $u_{pulsation}=A=3 c_s$ for Runs 04 - 20. Starting from a supersonic flow ensures a pulsation driven outflow without accretion. That leads to driving radial velocities higher than the observed photospheric radial velocities. \citet{Gillet.Fokin:14} found that even though shocks do form at the photospheric level for pulsating stars, in the case of classical Cepheids, the absence of a hydrogen emission lines indicate that those shocks have low Mach numbers. Thus in our simulations 04 - 20 the shocks already start with a Mach number M$=3$, which might be on the high side compared to self-consistent low atmosphere models \citep{Gillet.Fokin:14}. 
{We also ran simulations 01 - 03 using the observed radial velocity. The terminal speed for these cases is 40 - 80~km s$^{-1}$ and of the same order of magnitude as observed by \citet{Matthews.etal:12}, but with two orders of magnitude lower mass-loss rates than e.g.~Run 04, which reaches terminal speeds of the order of 200~km s$^{-1}$, i.e. one order of magnitude higher speed than observed by \citet{Matthews.etal:12}. It is thus difficult to reach the observed range of mass-loss rates with very low terminal speeds. Moreover, since in this study we focus mainly on exploring the best case scenario for X-ray production by the shock wave heating mechanism we performed the majority of our simulations with an already supersonic flow and self-consistently solved for the evolution of the shock dynamics of the system.}

\paragraph{Radiation pressure}
The simulations presented in this paper correspond to outflow purely driven by pulsations, having ignored any radiation pressure. Since radiation pressure is the accepted driver of winds in hot high-mass O-B stars \citep{Castor.eta:75}, it is worthwhile assessing its likely contribution as a source-term in the momentum conservation equation. We can evaluate its importance as a stellar wind driver by comparing it to the gravitational force. {Even though stellar winds in hot stars are line-driven, we can estimate the relative importance of the radiation pressure in Cepheids by comparing their continuum luminosity with the Eddington luminosity.} The luminosity of $\delta$~Cep is 2,000 L$_\odot$, which is much lower (by a factor 80) than the Eddington Luminosity for $\delta$~Cep, which is 160,000 L$_\odot$. Thus, the continuum radiation produces a force 80 times weaker than the gravitational pull applied to a parcel of fluid and thus it can be ignored and a purely pulsation-driven outflow is a realistic approximation in the case of $\delta$~Cep \citep{Castor.eta:75}. 

\subsection{Characteristic Time-scales}\label{sec:times}

\begin{figure*}[htbp]
 \begin{center}
 \includegraphics[width=0.8\textwidth]{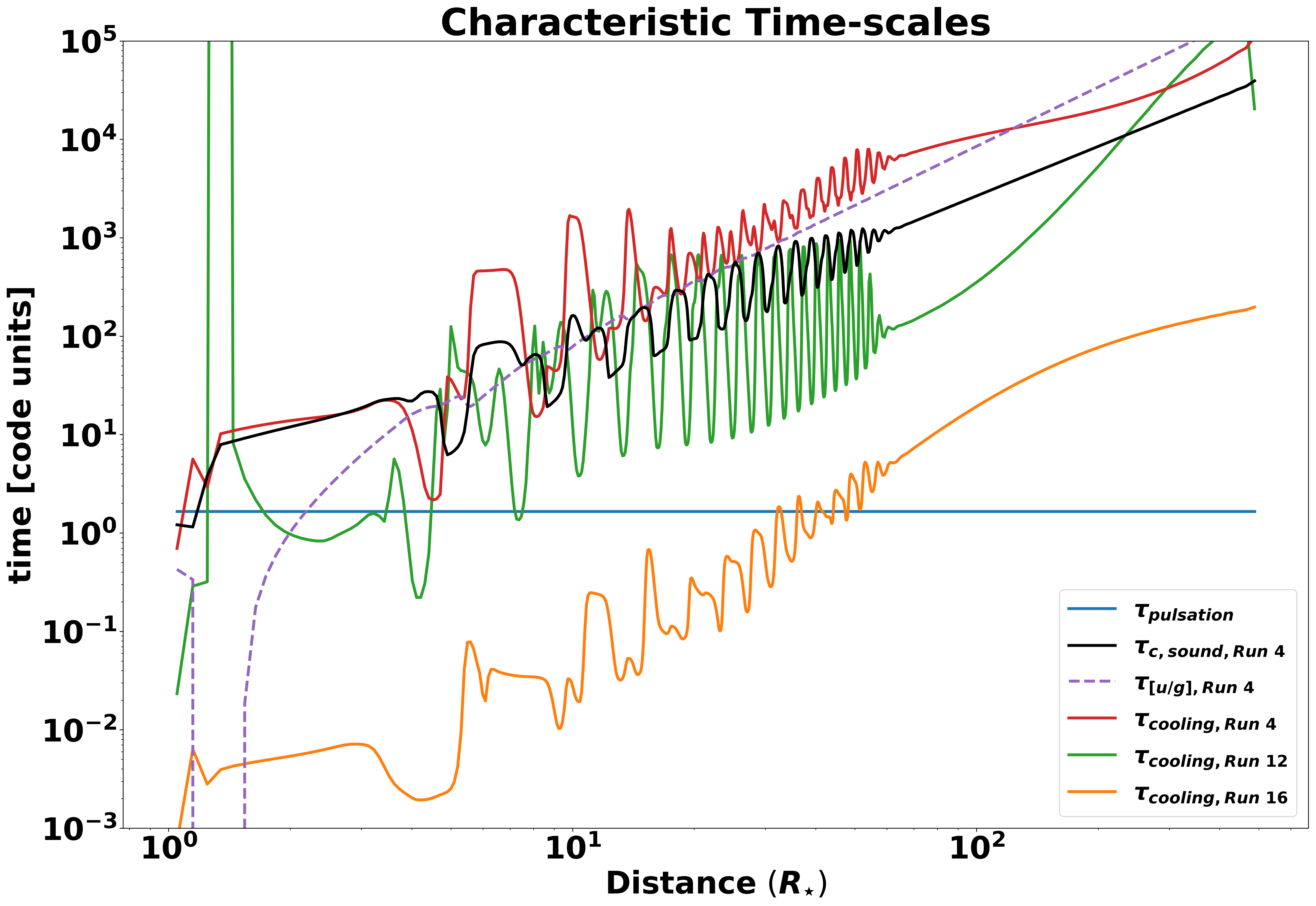}
  \includegraphics[width=0.8\textwidth,trim={0cm 0cm 0cm 0cm},clip]{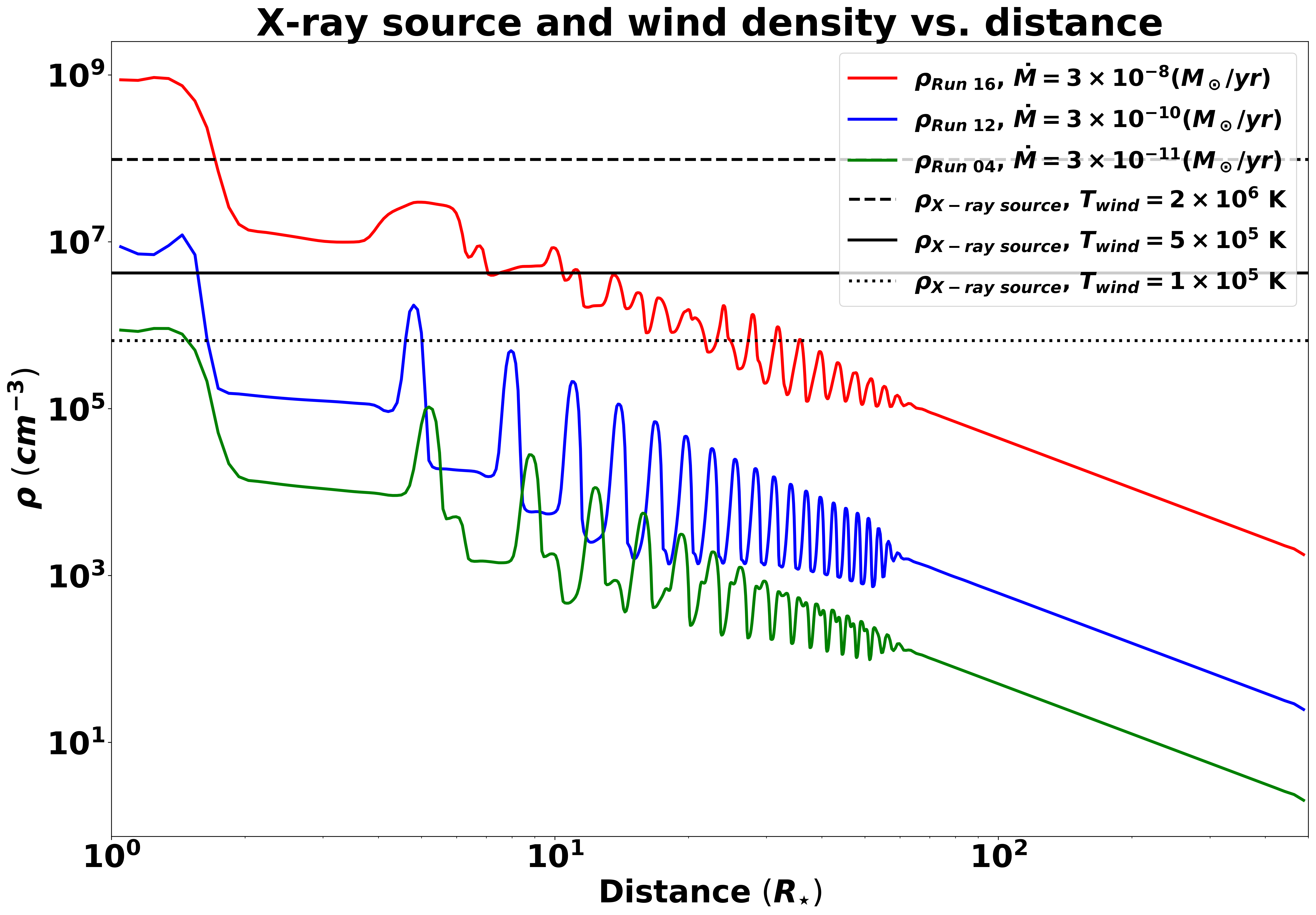}
  \caption{\textit{Top}: Characteristic time-scales for different quantities relevant to the simulations as a function of radial distance from the star. \textit{Bottom}: The density of the X-ray source assuming the observed X-ray decay for $\delta$~Cep is due to radiative cooling for two different temperatures of the shocked wind (horizontal lines), and the wind densities based on a typical terminal speed of 200 km s$^{-1}$ and for three mass-loss rates corresponding to Runs 04, 12, and 16. The cross over of these curves with the radiative cooling lines indicates the X-ray source location. The higher the mass-loss rate the further away the X-ray source is located.}
  \label{fig:times}
  \end{center}
  \end{figure*}

The dynamics in the problem is governed by several characteristic time-scales, namely the sound crossing time $[r/c_s]$, the pulsation period $[1/\omega]$, a dynamic time, i.e. the time-scale of the outflow to accelerate and get out of the gravity well of the star $[u/g]$, and finally the cooling time, i.e. the time it takes for the plasma to cool down or diffuse a specific compression. To check which time scales control the solution under different conditions, all characteristic time scales are compared in Fig.~\ref{fig:times}. For the cooling time, we use the analytic form of the cooling function obtained from the polynomial fit shown in Fig.~\ref{fig:lambda}. The shortest time scale is the one that controls the dynamics, as more change takes place due to that quantity in the simulation at a given time.
  
  For Run 04, close to the Cepheid the shortest and thus most dominant time-scale for the dynamics is the gravitational dynamic time-scale, and this is why we see a slow inflow in that region close to $r=1R_\star$. The pulsation time stays constant and it is the next shortest close to the star. From a distance $r=2R_\star$, the dominant phenomenon is the pulsation, which is verified by the fact that the X-ray emission peaks at that distance. Both the cooling time and the sound crossing time are similar up to $r=4R_\star$, and further away the sound-crossing time becomes the most relevant for the dynamics of the simulations. This means that the information about the state of the fluid is then propagated outwards without significant changes.
  
  For Run 12, which has ten times larger base density compared to Run 04, things are different. More specifically, for this case the cooling time is shorter than the sound crossing time for the largest part of the simulation domain, from $r=4R_\star$ onward. Thus we see that the cooling becomes a dominant effect with increasing density. This is also confirmed by the results of Run 16, which has a base density of $n_{e,0} = 10^9$ cm$^{-3}$, i.e. 1,000 times larger than Run 04. Even though Run 16 does not include any cooling, it can help us calculate the corresponding cooling time at each point from the simulation density and temperature. The top panel of Fig.~\ref{fig:times} shows that, apart from the region very close to the star, the cooling time dominates the entire region up to almost $100 R_\star$, which essentially means that it dominates in the entire computational domain, since from there on the solution is just propagated outwards.  
 
 \subsection{X-ray source Density}\label{sec:source}
  The bottom panel of Fig.~\ref{fig:times} shows an estimation of the density of the observed X-ray source for the two temperatures we have used for our simulations, $5\times10^5$ K and $2\times10^6$ K and a lower one, $1\times10^5$ K, and an approximate cooling time lasting for half a day as determined from the decay time of the X-ray peak in $\delta$~Cep. The density of the X-ray source is given by Eq.~\ref{eq:lambda}. {
  The figure also shows the density with distance for Runs 04, 12, 16 in red, blue and green colors, respectively. The intersection of the simulation curves with the observationally inferred lines gives the location of the emitted X-rays. }

  {From Fig.~\ref{fig:times}, we conclude that the range of most probable mass-loss rates is in the range $\approx 3\times 10^{-10}$--$3\times10^{-8}$ $M_\odot$ yr$^{-1}$. This is due to the fact that the density profiles corresponding to this range of mass-loss rates intersect the inferred X-ray source densities corresponding to the hot coronal plasma assumed in this study at acceptable distances. In general, the higher the base density assumed and the lower the atmospheric temperature, the further out are the X-rays expected to be emitted from.}  
  
  {More specifically, the red density curve corresponding to Run 16 with a mass-loss rate of $3\times10^{-8}$ $M_\odot$ yr$^{-1}$, intersects all three possible X-ray source densities corresponding to atmospheric temperatures $1\times10^5$, $5\times10^5$ K and $2\times10^6$ at distances close to $r=30$ , $r=10$ and $r=2$ $R_\star$. The $r=2$ $R_\star$ distance is the position of the shock where the X-ray emission peaks during the pulsation cycle.}
 
{The blue density profile corresponding to Run 12 and mass-loss rate $3\times10^{-10}$ $M_\odot$ yr$^{-1}$ intersects those of the X-ray source for both temperatures $5\times10^5$ K and $1\times10^5$ K at distances $r=2R_\star$, and $r=2R_\star$ and $r=5R_\star$, respectively. The green curve corresponding to Run 4 and mass-loss rate $3\times10^{-11}$ $M_\odot$ yr$^{-1}$ intersects only the density of the emitting plasma which corresponds to $1\times10^5$~K at distance $r=2R_\star$.
 A higher mass-loss rate, e.g. $10^{-6}$ $M_\odot$ yr$^{-1}$, would put the location of the X-rays at distances between $r=20R_\star$ and a few hundred $R_\star$ for the atmospheric temperature range considered here. Any distance larger than ten stellar radii seems very unlikely for the production of the observed thermal X-rays, since it is very difficult to achieve high enough temperatures and densities. Thus, in conclusion, the mass-loss rates that are acceptable considering a hot coronal plasma are roughly between $10^{-10}$ and $10^{-8}$ $M_\odot$ yr$^{-1}$.}

\subsection{Mass-loss rate}
Mass-loss rates in classical Cepheids are still not very well understood.  A wide range of estimated values have  been obtained from a variety of  approaches, ranging from $10^{-10}$ to $10^{-6}$ $M_\odot$ yr$^{-1}$ \citep[see, e.g.][]{Neilson.Lester:08,Marengo.etal:10b,Neilson.etal:11,Matthews.etal:12}. Both theoretical models and observational techniques are based on a number of strong assumptions that can heavily impact the mass-loss rate estimate.

The models presented here have fairly low mass-loss rates in the range $10^{-13}$-$10^{-8}$ $M_\odot$ yr$^{-1}$. From the previous Section~\ref{sec:source}, we argued that the mass-loss rates that are consistent with the X-ray observations are between $10^{-10}$ and $10^{-8}$ $M_\odot$ yr$^{-1}$. On the high end, the largest mass-loss rate was achieved for Run 16, which corresponds to a base density of $n_{e,0} = 10^9$ cm$^{-3}$, but ignores any cooling. Simulations 06 and 11 are then the most probable ones since they produce a mass-loss rate close to the lower limit of the acceptable mass-loss range. Both simulations correspond to a coronal temperature of $2\times10^6$ K, which corresponds to the solar coronal temperature. Run 06 produces a mass-loss rate of $5.55\times10^{-11}$ $M_\odot$ yr$^{-1}$ and a terminal speed of 580~km~s$^{-1}$, while Run 12 that has a 10 times larger density, produces a mass-loss rate of $2.67\times10^{-10}$ $M_\odot$ yr$^{-1}$ and a terminal speed of 170 km s$^{-1}$. At this point, it is worth mentioning that \citet{Matthews.etal:12} measured a terminal velocity of about 20 km~s$^{-1}$.

\subsection{X-ray emission}
We ran more than a dozen simulations of pulsation driven shocks in the corona of $\delta$~Cep. In our simulations, we were able to reproduce a wide range of peak X-ray luminosities. Our results indicate that, when in a pulsation driven stellar atmosphere, shock waves that drive the outflow are able to produce the luminosities observed at the X-ray emission peak phase. In all the cases, the emission was phase-dependent and it varied by 2 - 3 orders of magnitude from the quiescent to the ``flaring" phase. This suggests a \textit{two component emission} from Cepheid pulsating stars, with (a) shock waves being able to reproduce the phase dependent variable emission, and (b) a separate quiescent process being the dominant emission mechanism for the remaining time during a pulsation period. The idea of a two component emission is supported by the fact that non-pulsating supergiants have the same level of quiescent X-ray emission as Cepheids, as shown in \citet{Ayres:17} and noted earlier.

Fig.~\ref{fig:lightcurve} shows the X-ray light-curves for all the simulations of this study. Several interesting trends are evident. In the top left panel, we show the light-curves for Runs 01--03. Run 01 produces the strongest peak of that subset of simulations, with Run 03 producing the second largest peak, but quite similar to run 02. Run 01 does not include any cooling, and this is why larger densities and temperatures are forming that produce the strong peak, whereas Runs 02 and 03 include cooling.

In the top right panel, we show the X-ray light curves for Runs 04--06. Run 06 reaches the highest peak, $1.5\times10^{-15}$ erg cm$^{-2}$ s$^{-1}$, while both Run 04 and Run 05 only reach $0.5\times10^{-15}$ erg cm$^{-2}$ s$^{-1}$. Run 05 corresponds to a Cepheid with half the pulsation period of $\delta$~Cep and has thus two peaks during one full pulsation cycle of $\delta$~Cep.  Even Run 06 is not as X-ray luminous as the observations of $\delta$~Cep indicate, which reached a luminosity more than a factor 10 larger, $20\times10^{-15}$ erg cm$^{-2}$ s$^{-1}$.

The middle left panel shows the X-ray light curves produced when two shocks are generated in one pulsation cycle with different frequencies, i.e. $\omega$ and $2\omega$, and different wave amplitudes. Runs 07 and 09 reach a high peak in the X-rays of $\sim 1.12\times10^{-15}$ erg cm$^{-2}$ s$^{-1}$, while Runs 08 and 10 reach $\sim 6\times10^{-16}$ erg cm$^{-2}$ s$^{-1}$. The middle right panel corresponds to Runs 11 and 12, which are similar to Runs 06 and 04, respectively, but with 10 times larger densities. Run 11 reaches a peak flux of $1.34\times10^{-13}$ erg cm$^{-2}$ s$^{-1}$ and Run 12 a much lower peak of $\sim 5\times10^{-17}$ erg cm$^{-2}$ s$^{-1}$.

In the bottom left panel, we show simulations 13 and 14, which correspond to Cepheids with longer periods than $\delta$~Cep by a factor of 2 and 4, respectively, i.e. $\sim 11$ d and $\sim 21$ d. We see that the longer the period the weaker the X-ray peak luminosity, which is expected since a faster pulsation would produce stronger shocks by giving the atmosphere less time to react to the perturbations. 

Finally, the bottom right panel shows the ideal cases without any cooling for Runs 16 - 19, with a factor 10 difference in the base density for each case and with Run 16 having the largest base density in the study, $n_{e,0} = 10^9$ cm$^{-3}$. The peak X-ray luminosities reached range from $\sim 6\times10^{-16}$ erg cm$^{-2}$ s$^{-1}$ for Run 19 to $\sim 1.4\times10^{-12}$ erg cm$^{-2}$ s$^{-1}$ for Run 16.

We thus conclude that the simulation that is closest to the observed X-ray peak flux is Run 12. If we wanted to reproduce the exact luminosity observed we should have run a simulation with base density four times larger than Run 04, i.e. $n_{e,0} = 4\times10^6$ cm$^{-3}$, since the emission has a square density dependence, $L_X \propto n_e^2$. 

{In Fig.~\ref{fig:phasecool}, we show the evolution by showcasing four snapshots of the hydrodynamic properties of Run 04 throughout one pulsation period. Both the density and the temperature vary by about one order of magnitude during the latest pulsation cycle, which corresponds to the newly emerged shock located closest to the stellar photosphere. This corroborates the results presented in Table~\ref{tab:sims} and Fig.~\ref{fig:lightcurve} that show the luminosity varying by about three orders of magnitude throughout a single pulsation cycle. Around phase 0.15 the light curve reaches its peak due to the shock. The shock is located at a distance of 2 $R_\star$ during the peak emission. During the rest of the pulsation cycle the shock wave emission drops quickly below the quiescent levels and the quiescent emission takes over. This phase-dependent behavior indicates that the emission is coming from a region below 10 $R_\star$ and is due to only the newly emerging shock. In this study, we ignore the effect of the absorption in the synthetic luminosity. In reality, however, the resulting light curve should emerge from a competition between the distance from the star and the absorption that each emitting parcel of fluid in the escaping atmosphere experiences. We note at this point, that the main contributing factor in the emission is the most recently emerged shock, since the luminosity has a strong inverse power law dependence on the shock distance from the star $L \propto \rho^2 \propto r^{-4}$. As Fig.~\ref{fig:phasecool} indicates, after one pulsation period each shock travels about 5 $R_\star$ for Run 04 and thus all previous shocks contribute 625 times less than the more recently emerged shock ($5^{-4}=1/625$).}

\subsection{Can Shocks Explain the X-ray Variability of Classical Cepheids?}
 
To date, there are only two classical Cepheids in which strong X-ray variability has been seen, namely $\delta$~Cep and $\beta$ Dor. 


\paragraph{$\beta$~Dor} The X-ray luminosity of $\beta$~Dor is  $\log{L_X} = 29$ erg s$^{-1}$  for a distance of 320~pc \citep{Engle.etal:09}. It has a mass of $M=6.5M_\odot$, a radius of  $R=67.8 R_\odot$ and an effective temperature of 5,445 K \citep{Turner:10}. It is  moving away from the Earth with a radial velocity of 9km s$^{-1}$, and pulsates with an amplitude similar to $\delta$~Cep \citep{Taylor.Booth:98}. 
The luminosity of $\beta$~Dor is 3,200 L$_\odot$, i.e. much smaller than its Eddington luminosity of 210,000 L$_\odot$, thus radiation pressure is negligible, similar to $\delta$~Cep. $\beta$~Dor has a similar escape velocity to  $\delta$~Cep---192~km s$^{-1}$ at the photosphere---but a pulsation period of $T_p=9.842$~d, which is almost double that of $\delta$~Cep
\citep{Klagyivik.Szabados:09}.

{The most representative run for $\beta$~Dor would be Run 13 that has double the pulsation period of $\delta$~Cep. In order to constrain the base density, we need again an appropriate photospheric model \citep{Lester.Neilson:08} and a better phase coverage in future observations. That will allow us to tune the base density to reproduce the observed luminosity.}

Other interesting cases include $\zeta$~Gem and Polaris ($\alpha$ Ursae Minoris). $\zeta$~Gem is quite similar to $\beta$~Dor in stellar parameters \citep[$T_p=10.15$~d, $M=7.7M_\odot$, $R=65.24 R_\odot$, $L=2900L_\odot$ $\log{L_X}$$ = 29.1$ erg s$^{-1}$][]{Tetzlaff.etal:11,Majaess.etal:12,Breitfelder.etal:16}, and has a tentative X-ray enhancement detection by \emph{EINSTEIN} almost 40 years ago \citep{BohmVitense.Parsons:83}. No X-ray variations have been observed in the short-period Cepheid Polaris \citep[$T_p=3.969$~d, $M=5.9M_\odot$, $R=53.6 R_\odot$, $L=1260 L_\odot$ $\log{L_X}$ $=$ 28.9 erg s$^{-1}$][]{Fernie.etal:95,Engle:15PhDT,Bond.etal:18}.  Both Polaris and $\zeta$ Gem have circumstellar envelopes  \citep{Merand.etal:06, Hocde.etal:20}. 

{From our simulations, $\zeta$~Gem would be best described by Run 13, as it has twice as long a period compared to $\delta$~Cep, while the Polaris case would be represented better by Run 04 or Run 05, as it has a comparable, but shorter period than $\delta$~Cep. Subsequently, we expect both stars to produce X-ray light curves with lower peaks by 30\% and 20\%, respectively, (the difference between X-ray peak emission in Run 13 and Run 05 compared to Run 04), compared to $\delta$~Cep if all other parameters are the same. These lower peaks in combination with their higher quiescent X-ray luminosities compared to $\delta$~Cep, which has $\log{L_X}$ $=$ 28.9 erg s$^{-1}$, could explain why no X-ray variation has been observed to date for these two stars.}



\subsection{Magnetic Reconnection}
Another possibility for the X-ray enhancements observed in Cepheids is that it is produced when magnetic field lines reconnect during the beginning of the collapse of the atmosphere, triggering a phase-dependent \emph{flare}. In the Sun, magnetic reconnection accelerates particles which then interact with the solar atmosphere and emit radiation (flare) throughout the entire electromagnetic spectrum. In the solar flaring scenario, UV and optical emission come from the lower atmosphere impacted by accelerated particles, while soft X-rays are produced higher up by hot plasma evaporated by the energetic particle impact. This scenario could explain the multi-frequency observations of $\delta$~Cep shown in Fig.~1 of \citet{Engle.etal:14}. {More specifically, starting from minimum radius as the expansion progresses, reconnection events could be triggered and combined with an increase in the chromospheric density due to the shock propagation, a corresponding UV enhancement could be induced locally. Gradually, as the expansion reaches the maximum radius phase, an X-ray enhancement could be provoked at higher altitudes as particles accelerated by reconnection events start colliding with the collapsing photosphere. Multi-wavelength flaring emission and a clear connection between UV and X-ray radiation is also seen in non-pulsating main sequence stars as a result of the stratification of the stellar atmospheres and the location of the reconnection at the apex of coronal loops \citep[see, e.g.][]{Neupert:68,Gudel:04, Osten.etal:04}.}

\noindent Magnetic fields are the main driver for activity in Sun-like and active stars \citep[see e.g.][]{Wright.etal:17}. Cepheids have outer convection zones and as such they are expected to produce a continuous low level X-ray emission, since yellow giants that are less luminous are known to emit in X-rays. The only Cepheid with a definite magnetic field detection is $\eta$~Aql, while tenuous magnetic field estimations for $\delta$ Cep are in the order of 1 G on average \citep{Grunhut.etal:10}. If the solar paradigm where the average magnetic field strength is about 100~G is any indication, then we could have regions with magnetic field strengths up to 30~G in active regions \citep{Okamoto.Sakurai:18}. Indeed, this tentative magnetic field strength has been confirmed by \citet{Auriere.eta:15} who observed the magnetic fields of about 30 active single G-K giants with magnetic fields in the range 1-40 G.

The magnetic energy conversion rate can be quantified using the scaling law of \citet{Mozer.Hull:10}, which depends on magnetic field, $B$, and particle number density, $n_0$, and is written as $f  \propto B^3 / \sqrt{n_0}$ per unit area. 
The plasma density of the emitting volume is estimated as $\sim 2\times 10^9$~cm$^{-3}$ from the decay time of the X-ray enhancement \citep{Engle.etal:17}. For this density and a typical magnetic field of the order of 1~G, the Cepheid reconnection rate is $f_{\delta\ Cep}=4.5\times 10^4$~erg~cm$^{-2}$~s$^{-1}$. A coverage of 2\%\ of the Cepheid surface is needed to explain the observed X-ray flux of  $10^{29}$~erg~s$^{-1}$. We will examine this possibility in future work.

\section{Summary and Conclusions}
\label{sec:conclusions}

Shocks waves are ubiquitous in the universe as X-ray sources. In this work, we explored the scenario of a pulsation-driven shock wave mechanism to create the observed \emph{Chandra} and \emph{XMM-Netwon} X-ray variability in classical Cepheids. We ran a set of about twenty simulations with a wide range of parameters in order to better understand the phenomenon for short period Cepheids, using $\delta$~Cep as the working example. 
We explored scenarios with different base densities, atmospheric temperatures, cooling functions, single and multiple driving waves with different amplitudes, different pulsation periods and simple analytic sinusoidal wave forms of proper Fourier decomposition to drive the pulsation from the radial velocity profile at the inner boundary of the simulation.

We were able to reproduce a shock-driven outflow, with small inflows very close to the star, as in previous studies. The terminal speeds of the outflows range from 40 to $\sim$ 600 km s$^{-1}$. Guided by the observations, which indicate the presence of hot million kelvin X-ray emitting plasma, we set the atmospheric temperature to the coronal values $5\times10^5$~K and $2\times10^6$~K. We covered a wide range of base density from $10^6$ cm$^{-3}$ to $10^9$ cm$^{-3}$, and confirmed that if the shocks are strong enough they completely dominate the atmospheric dynamics and the initial radial profile of the density converges to that of a steady outflow, $\rho = \rho_0 (R_\star/r)^2$. We examined the role of the cooling in the shock wave scenario, by running both ideal simulations without any cooling processes and including the tabulated cooling function in \emph{PLUTO}. The cooling is an essential ingredient for capturing the shock dynamics properly and dominates the simulation dynamics for large densities, as the cooling time becomes shorter than all other time-scales.

Our results span a wide range of mass-loss rates, $2\times10^{-13}$--$3\times10^{-8}$ $M_\odot$ yr$^{-1}$ and peak X-ray luminosities $5\times10^{-17}$--$1.4\times10^{-12}$ erg cm$^{-2}$ s$^{-1}$. Furthermore, we showed that, when the appropriate conditions are met, shock waves can reproduce the observed characteristic X-ray behavior. More specifically, we found that a simulation similar to Run 6 with 4 times larger density would be the most appropriate to capture the peak X-ray luminosity of $\delta$~Cep, $20\times10^{-15}$ erg cm$^{-2}$ s$^{-1}$, and a mass-loss rate of about $\sim10^{-10}$ $M_\odot$ yr$^{-1}$. All of our simulations produced a phase-dependent emission increase. {In order to properly reproduce the phase of the X-ray peak, we would need to include higher order harmonics with degrees $m>2$, as done for Runs 01 - 03. The phase-dependent emission increase is due to the most recently emerged shock wave and all the previous shock waves are only minority contributors to the total luminosity due to the sharp decay of the luminosity with density and distance from the star as $L\propto \rho^2 \propto r^{-4}$.}  

{More specifically, for all our simulations, we showed that the X-rays are emitted from distances close to the star. For the basic case Run 04, we showed that the peak emission is coming from a distance of about $2 R_\star$. In most simulations that were driven at the inner boundary by simple sinusoidal functions of either the first term or two first terms of the harmonic expansion, the shock wave X-ray emission enhancement appeared to take place at phases 0 - 0.4. In all cases, the X-ray variability within a pulsation cycle was very significant, varying by orders of magnitude. This indicates a two-component scenario comprising a quiescent emission component that dominates during most phases and a shock-produced X-ray emission that dominates within the phase interval 0.2 - 0.6 for the Fourier cases (Runs 01 - 03) and 0 - 0.4 for all other cases (Runs 04 - 20). The difference in the phase of the shock-wave emission enhancement between Runs 04 - 20 and the observations, probably comes from the fact that the former ignore the higher order harmonics of the velocity decomposition with degree $m>2$, that can cause a shift in the timing of the emission peak. When we include five, instead of one or two, terms of the harmonic expansions for the Fourier cases (see Fig.~\ref{fig:lightcurve}) the phases of the X-ray emission enhancement of Runs 01 - 03 agree with the observations that show an enhancement during phases 0.4 - 0.6 \citep{Engle.etal:14}.}


$\delta$~Cep is not the only Cepheid with observed variability. We commented on the expected X-ray behavior from other classical Cepheids with pulsation periods similar to, or larger than, that of  $\delta$~Cep. 
More X-ray observations are necessary to better constrain the energetic properties of these classical Cepheids. 



\acknowledgments
\noindent S.P.M. was supported by \textit{Chandra} Modeling CHANDRA Observations of Cepheid Activity theory grant number 21200202. J.J.D. was funded by NASA contract NAS8-03060 to the \textit{Chandra} X-ray Center and thanks the director, Belinda Wilkes, for continuing advice and support. Support was provided to N.R.E. by the {\it Chandra} X-ray Center NASA Contract NAS8-03060 and by NASA through \textit{Chandra} grants to NRE GO7-18013X and GO8-19008X.  J.G. acknowledges participation in the MESA 2019 summer school, supported in part by the National Science Foundation under Grant No. NSF ACI-1663688. Support was provided to J.A.Z. by the {\it Chandra} X-ray Center NASA Contract NAS8-03060. H.R.N is grateful for funding from a Discovery Grant from the Natural Sciences and Engineering Research Council of Canada.

\software{\textit{PLUTO} \citep{Mignone:14}}

\bibliographystyle{yahapj}
\bibliography{references}

\appendix
{
\section{Complete simulation results} \label{sec:appendix}
Here we provide a description of each separate simulation setup with the most notable results in each case.
\paragraph{{\bf Run 01}} 
First, we used the exact Fourier decomposition corresponding to the photospheric radial velocity observed to drive the pulsation, but the steady state solution resulted in accretion rather than an outflow. This is due to the fact that the initial velocity was too low to drive an outflow. The initial velocity needs to be larger than the local escape speeds, which at $r=R_{\delta-Cep}$ is $219$ km s$^{-1}$. For that reason we increased the driving velocity by a factor 10, as shown in Table~\ref{tab:sims}. The physical argument behind this is that the observed radial velocity is at the photospheric level, whereas our simulation only treats the coronal plasma, at higher altitudes, where the density is lower, and where an initial outgoing perturbation is expected to accelerate due to the density stratification and the corresponding pressure gradient.

The resulting steady state solution is shown in the top panel of Fig.~\ref{fig:1000Puls}. The temperature in the left panel is decreasing outwards as a power law, similar to the density profile shown in the right panel. More specifically, the steady state convergent solution creates a density profile that drops as $\rho\propto r^{-2}$ in the gravitational field of the Cepheid. 
The temperature profile is then expected to be $T\propto P/\rho=\rho^{\gamma}/\rho=\rho^{\gamma-1}=r^{-4/3}$. And this is indeed the power law that the steady solution follows. The velocity shows some initial oscillations due to the pulsation driver at the inner boundary, which gradually decay, and a terminal outgoing speed of about 80 km s$^{-1}$ is reached. 

\paragraph{{\bf Run 02}} 
For this simulation, we use the exact same setup as for Run 01, with the difference that now we include a radiative cooling function given by the polynomial fit presented in section \ref{sec:methods}.
We included heating as described above and obtained solutions consistent with the method of decreasing the polytropic index. The heating is not very relevant as a time-scale since $\tau_h=R_\star/u_{flow}\sim 350$~ks, but also for the driving of the outflow, since probably it is fully shock wave driven. Thus, we do not decrease the polytropic index or include an explicit heating analytically.
The cooling works well and the time-step is similar to the cases without cooling. 
In the simulations with cooling included, we need an even stronger shock to keep seeing the same bumps in the temperature and density profiles, as for Run 01, when cooling was excluded. 
In this case, the profiles of temperature, and density are much more disturbed, with major cooling taking place around $r=2 R_\star$ and the temperature dropping dramatically.

\paragraph{{\bf Run 03: Tabulated Cooling}} 
For this simulation, we use the same setup as Run 02, only that now the cooling is given by a tabulated form as discussed in \citet{Mignone:14} and shown by the blue curve in Fig.~\ref{fig:lambda}. {The polynomial fit follows closely the more accurate tabulated form of cooling as shown in Fig.~\ref{fig:lambda}. The differences between Run 02 and Run 03 as presented in Fig.~\ref{fig:1000Puls} can be explained by the fact that the hydrodynamic equations are non-linear and small changes in the energy equation can result in noticeable differences.} The tabulated cooling function is more accurate with a similar computational cost as the polynomial function, so we decided to use the tabulated cooling for all the cases hereafter.

\paragraph{{\bf Run 04}} The same setup as Run 03 is used, with the exception that for the velocity we now have a simple sinusoidal function with amplitude $A = 3 c_s$ and a frequency corresponding to the pulsation frequency $\omega = 2\pi/T_p$. This parametric form of the outflow ensures a large enough speed that can escape the gravitational attraction of the star. For this case, the local sound speed at $r=R_\star$ is about 100~km s$^{-1}$ and the pulsation amplitude is set to 3 times that, i.e. 300 km s$^{-1}$, which is larger than the local escape speed and thus the material escapes and a steady state outflowing wind solution is formed. The terminal speed of the converged solution is 200~km~s$^{-1}$, similar to the shock speeds suggested by \citet{Engle.etal:17}.

This solution is significantly different compared to the Fourier decomposition cases, in that the driving velocity perturbation is now larger. The perturbation enforced at the inner boundary is then creating characteristic oscillations in the profiles of all the quantities that gradually decay at a distance of about 100~$R_\star$.

\paragraph{{\bf Run 05}} For this run, we use the same setup as Run 04, but with twice as high driving frequency, i.e. $2\omega$. Due to this higher frequency, the first peak appears earlier compared to Run 04 in all profiles. The oscillations are also fading faster compared to the lower frequency case, but the terminal speed and average values for temperature and density agree for both solutions, as the velocity amplitude and all other physical quantities remain the same. The obtained light curve shows two peaks close to the two edges of the numerical domain, with the luminosity varying by about 3 orders of magnitude, as shown in Fig.~\ref{fig:lightcurve}.

\paragraph{{\bf Run 06}} For this run, we use the same setup as Run 04, but with a background temperature of $T=2\times10^6$K. This leads to a distinctly different solution. The terminal speed is now significantly different, 600 km s$^{-1}$, i.e. a factor 3 higher, compared to Run 04. The first peaks in all profiles appear later on, as for a warmer plasma the characteristic sound speed is larger and the perturbations propagate faster. A higher average temperature and density are reached at larger distance, as shown in the second row of Fig.~\ref{fig:1000Puls}. The oscillations have larger amplitudes initially, but fade fast, at a similar distance as for Run 04. The peak of the light curve for Run 06 is visibly higher than Run 04 and Run 05, as shown at the top right panel of Fig.~\ref{fig:lightcurve}.

\paragraph{{\bf Run 07}} We use the same setup as for Run 04, but this time instead of a simple sinusoidal function for the velocity, we use the combination of two sinusoidal waves with the same amplitude, $A$, and two different frequencies $\omega$ and $2\omega$. A coupling between these two frequencies seems to be taking place, with an early smaller peak and a large second peak corresponding to the global maximum in all quantities. The terminal speed reached is now double that compared to Run 04 and Run 05, i.e. 400 km s$^{-1}$, as the velocity perturbation amplitude has now doubled.

\paragraph{{\bf Run 08}} Here we use the same setup as Run 07, with the only difference being that the two sinusoidal components of the velocity have amplitudes equal to half of the initial amplitude, i.e. $A/2$. The results are similar to Run 07, with the peaks appearing earlier and the terminal speed being 110 km s$^{-1}$, as a result of the coupling between the two modes with different frequencies. The average temperature is lower compared to Run 07.

\paragraph{{\bf Run 09}} Here we use the same setup as Run 07, with the first wave of frequency $\omega$ having an amplitude of $A$ and the second wave of frequency $2\omega$ having half of that amplitude $A/2$. The terminal speed is now 280 km s$^{-1}$ and the solution is very similar to those from Run 07 and Run 08, with the average temperature being between the temperatures found for those runs.

\paragraph{{\bf Run 10}} Here we use the same setup as Run 07, with the first wave of frequency $\omega$ having an amplitude of $A$ and the second wave of frequency $2\omega$ having a quarter of that amplitude $A/4$. The terminal speed is now 230 km s$^{-1}$ and the solution is otherwise very similar to Run 08.

\paragraph{{\bf Run 11}} For this run, we use the same setup as Run 06, but with a base density at the inner boundary 10 times higher. The terminal speed is 560 km s$^{-1}$ and the oscillation amplitudes are large, but the range of the shock region appears much narrower. The peaks appear at the same point as in Run 06, which is expected for plasmas with the same temperature and thus sound speeds.

\paragraph{{\bf Run 12}} For this run, we use the same setup as Run 11, but for a coronal temperature of 500,000K. The terminal speed is 270 km~s$^{-1}$, the perturbation amplitudes are lower and the peaks appear earlier. The temperature at distances less than 10~$R_\star$ are very low, reaching a few tens of K. The corresponding peak of the light curve in Fig.~\ref{fig:lightcurve} is significantly lower compared to the hot case of Run 11.

\paragraph{{\bf Run 13}} For this run, we use the same setup as Run 04, but for a pulsation period twice as large as that of $\delta$~Cep. Run 04 shows its peak earlier on, and the oscillations have smaller amplitudes, see Fig.~\ref{fig:1000Puls}. Furthermore,  the computed luminosity peak for Run 13 is lower and asymmetric compared to the symmetric and more intense light curve of Run 04.

\paragraph{{\bf Run 14}} For this run, we use the same setup as Run 04, but for a pulsation period four times as large as that of $\delta$~Cep. Run 13 shows its peak earlier on, and the oscillations have smaller amplitudes, see Fig.~\ref{fig:1000Puls}. Furthermore,  the computed luminosity peak for Run 14 is lower and more asymmetric compared to the more symmetric and more intense light curve of Run 13.

\paragraph{{\bf Run 15}} The setup of this run was based on Run 12 with a 10 times larger initial and base density, i.e. $n_e=10^8$ cm$^{-3}$. The cooling in this case is so significant and the cooling time-scale so small that the simulation becomes challenging and very computationally expensive. We had to include an extra higher resolution region close to the star where the high densities are located, with 2480 linear points within the radial distance range [1,10], 496 linear points within [10,50], and 100 logarithmic points in [50,500]. This helped the simulation run for more than 550 pulsation periods and propagate the initial perturbation to the outer boundary. The simulation, however eventually crashed due to the intensity of the cooling close to the star. The obtained results suffice to get a good idea of what the steady state solution looks like, so we chose to not persist to 1000 pulsations.

\paragraph{{\bf Runs 16 - 19}} Finally, we switched the cooling off and ran 4 cases based on Run 04 with the same density, 10 times, 100 times, and 1000 times larger base densities than that of Run 04, as shown in Table~\ref{tab:sims}. The results are very interesting. All terminal speeds are now 230 km s$^{-1}$, and the computed mass-loss rates are differing by exactly one order of magnitude between each case progressively. A similar trend is also followed in the X-ray light curves. Fig.~\ref{fig:1000Puls} shows the reason for this. The density and thus the pressure profiles are exactly the same, just differing by an order of magnitude, thus the temperature and the sound speed, that depend on the ratio $P/\rho$, are identical for all four cases. Subsequently, the terminal speed which depends on the initial perturbation and the sound speed are also identical.

\paragraph{{\bf Run 20}} In order to check how important the initial conditions and specifically the initial density profile is, we ran a last simulation based on Run 04, but with an initial density profile that drops as $\rho \propto (R_\star/r)^{4}$ instead. The steady state solution that the code converges to is exactly the same as for Run 04. Thus, we conclude that the shock dynamics completely take over and dominate the evolution of the solution completely wiping out the initial conditions. 
}
 
\end{document}